# Polymer-based Hybrid Integrated Photonic Devices for Silicon On-chip Modulation and Board-level Optical Interconnects

Xingyu Zhang, Amir Hosseini, *Member, IEEE*, Xiaohui Lin, Harish Subbaraman, *Member, IEEE*, and Ray T. Chen, *Fellow, IEEE*

***Abstract*—The accelerating increase in information traffic demands the expansion of optical access network systems that require cost reduction of optical and photonic components. Low cost, ease of fabrication, and integration capabilities of low optical-loss polymers make them attractive for photonic applications. In addition to passive wave-guiding components, electro-optic (EO) polymers consisting of a polymeric matrix doped with organic nonlinear chromophores have enabled wide-RF-bandwidth and low-power optical modulators. Beside board level passive and active optical components, compact on-chip modulators (a few 100µm to a few millimeters) have been made possible by hybrid integration of EO polymers onto the silicon platform.**

**This paper summarizes some of the recent progress in polymer based optical modulators and interconnects. A highly linear, broadband directional coupler modulator for use in analog optical links and compact, and low-power silicon/polymer hybrid slot photonic crystal waveguide modulators for on chip applications are presented. Recently, cost-effective roll-to-roll fabrication of electronic and photonic systems on flexible substrates has been gaining interest. A low-cost imprinted/ink-jet-printed Mach–Zehnder modulator and board-to-board optical interconnects using microlens integrated 45-degree mirror couplers compatible with the roll-to-roll fabrication platforms are also presented.**

## I. Introduction

Polymer active and passive optical components have shown potentials for enabling economically viable optical systems crucial for cost-sensitive applications, such as Fiber-To-The-Home (FTTH) networks [1]. The low-temperature process and wet-coating process of large areas suitable for almost all kinds of substrates are key requirements for mass production potentials. Polymer waveguide materials can be highly transparent, and the absorption loss can be made below 0.1dB/cm at all key communication wavelengths [2]. The refractive index of most polymers (1.5-1.7) is nearly matched to that of glass optical fibers (1.5-1.6) enabling small Fresnel reflection loss at the interfaces in butt-coupling configuration of waveguides and I/O fibers.

In addition to passive components, the Thermo-Optic (TO) and Electro-Optic (EO) effects in polymers have been used for the realization of active devices [3-5]. EO polymer based optical modulators offer several advantages over the mature lithium niobate ($LiNbO_3$) modulators due to the exclusive properties of polymer materials [2, 6-9]. EO polymers can be engineered to have very large EO coefficients, $\gamma_{33}$, which is advantageous for sub-volt half-wave switching voltage ($V_\pi$) [10-12]. For example, CDL1/PMMA, an EO polymer with $\gamma_{33}$=60pm/V, was used to achieve $V_\pi$=0.8V [10]. Another EO polymer with a very large $\gamma_{33}$=306pm/V was developed through controlled molecular self-assembly and lattice hardening [13]. In comparison, the $\gamma_{33}$ of $LiNbO_3$ is only about 30pm/V at 1300nm.

Based on the advantages of polymer materials, EO polymer modulators have shown great potentials for a variety of applications, such as telecommunication and digital communication [14, 15], analog-to-digital conversion [16], phased-array radar [17], and electromagnetic field sensing [18], etc. Excellent velocity matching between microwaves and optical waves can be achieved due to a close match between the refractive index of polymers at microwave and optical frequencies, enabling ultra-broad bandwidth operation. Additionally, the intrinsic relatively low dielectric constant of polymers (2.5-4) enables 50-ohm traveling wave electrodes to be easily achieved. In earlier work, a traveling wave polymer modulator with the bandwidth of 40 GHz was demonstrated by Teng [5]. Later on, a polymer modulator operating over 100GHz was developed by research groups at UCLA and USC [19]. Up until today, the highest frequency for a polymer modulator has been demonstrated to be up to 200GHz by Bell Laboratories [20]. Recently, polymer based modulators with high reliability have become commercially available [21, 22]. The polymer based optical modulators are more advantageous for broadband operation over other modulators based on gallium arsenide (GaAs), indium phosphide (InP) or silicon. For reference, 10GB/s and 40Gb/s nonreturn-to-zero and return-to-zero GaAs modualtors with $V_\pi$ less than 5V have been demonstrated [23, 24]. A transponder using a tunable InP-based MZI modulator with a maximum likelihood sequence estimation (MLSE) receiver has already been demonstrated for 10Gbit/s [25]. Silicon modulator based on carrier depletion of a *pn* diode embedded inside a silicon-on-insulator (SOI)

X. Zhang, X. Lin, and R. T. Chen are with the Microelectronic Research Center, Department of Electrical and Computer Engineering, University of Texas, Austin, TX 78758 USA (e-mail: xzhang@utexas.edu; raychen@uts.cc.utexas.edu).

A. Hosseini (amir.hosseini@omegaoptics.com) and H. Subbaraman are with Omega Optics, Inc., Austin, TX 78759 USA.

waveguide has been demonstrated to have a 3 dB bandwidth of ~30 GHz and can transmit data up to 40 Gbit/s [26].

Compared to the established technology platforms such as $LiNbO_3$, polymer materials show some advantages in processing. For example, polymers can be easily spin-coated onto almost any genre of materials. In addition, different from the difficult implementation of the domain inversion technique on $LiNibO_3$ [27], Δβ-reversal can be easily achieved by domain-inversion poling on EO polymers. Furthermore, the integration of high performance EO polymer films onto SOI wafers has enabled compact EO polymer modulators including EO polymer refilled silicon slot waveguides [28, 29] and slot photonic crystal waveguides (PCWs) [30, 31] for on-chip applications.

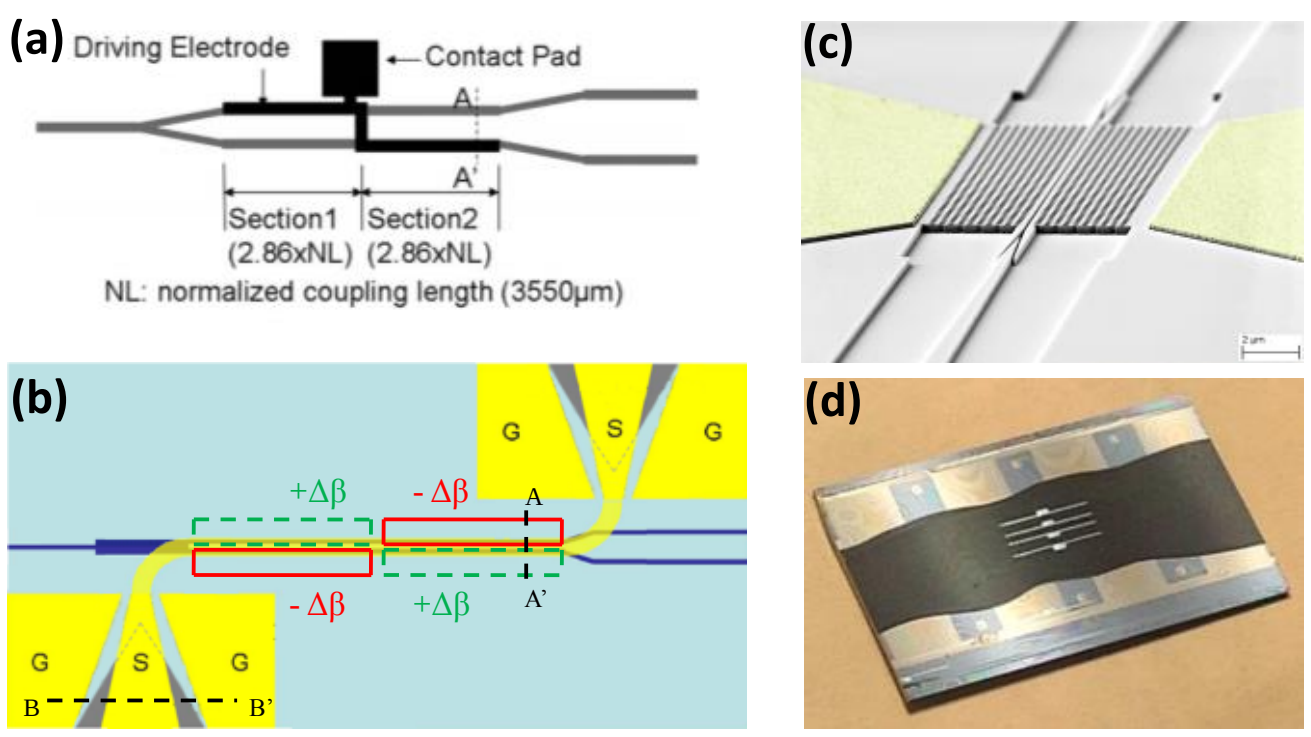

Fig. 1. (a) 2-domain YFDC modulator based on EO polymer, with lumped element driving electrode. (b) Traveling wave MMI-fed 2-domain directional coupler modulator based on EO polymer. The red solid lines and green dashed lines indicate the area of EO polymer poled in opposite directions, respectively. (c) EO polymer refilled silicon slot PCW MZI modulator. (d) MZI polymer modulator fabricated by UV imprinting and ink-jet printing.

In this paper, we present a few recent developments in EO polymer based modulators for on-chip and board-level applications (Fig. 1) in addition to passive and active ink-jet-printable devices for roll-to-roll fabrication platforms. The rest of the paper is organized as follows. A highly linear, broadband polymeric directional coupler modulator with spurious-free dynamic range (SFDR) of $119dB/Hz^{2/3}$ is presented in Section II. Existing commercial $LiNbO_3$ Mach-Zehnder interferometer (MZI) modulators with sinusoidal transfer functions suffer from intrinsic nonlinear distortions that limit their application in analog links. By applying the Δβ-reversal technique through domain-inversion, we show that the linearity of the device can be improved when the intermodulation distortions in two domains cancel each other out. Section III describes the first EO polymer refilled slot PCW based MZI modulator with a 308μm-long active region. The short device length is made possible by 1) slow-light effects of the slot PCW, and 2) concentration of high photon energy in the slot region. The slot PCW structure can be further optimized to improve the optical wavelength range of operation and the poling efficiency as described in Section IV. An effective in-device $\gamma_{33}$=1012pm/V and a voltage-length product of $V_\pi \times L$=0.345V mm were demonstrated in a band-engineered EO polymer refilled slot PCW modulator. In Section V, we present a low-cost EO polymer MZI modulator entirely fabricated by molding and ink-jet printing. Finally, low-cost board-to-board optical interconnects using couplers consisting of imprinted 45° mirrors and ink-jet printed microlenses are demonstrated to operate at over 10Gbps with bit error rate (BER) better than $10^{-9}$ in Section VI.

## II. Conventional Polymer Waveguide Modulator with High Linearity and Broad Bandwidth

Optical modulators in analog optical links are required to have high modulation efficiency, good linearity and large bandwidth. Existing commercial $LiNbO_3$ MZI modulators have intrinsic drawbacks in linearity to support high fidelity communication. When multiple tones of signals ($f_1$ and $f_2$) are simultaneously carried over a link, nonlinear intermodulation distortion signals are generated. The third-order intermodulation distortions (IMD3), which are the byproducts of the interaction between fundamental frequencies and harmonics and occur at ($2f_1$- $f_2$) and ($2f_2$- $f_1$), are considered the most troublesome among all the nonlinear distortions because they usually fall within the usable bandwidth of the system. The spurious free dynamic range (SFDR) is defined as the dynamic range between the smallest signal that can be detected in a system and the largest signal that can be introduced into the system without creating detectable distortions in the bandwidth of concern [32]. The SFDR of high frequency analog optical links is limited by the system noise and the nonlinearity of modulation process.

Bias-free Y-fed directional coupler (YFDC) modulators have been shown to provide better linear transfer functions compared to the sine-squared transfer curve of conventional MZI modulators [33, 34]. The device linearity can be further enhanced when the YFDC modulators are incorporated with the Δβ-reversal technique to suppress IMD3s [33, 35-38]. We previously demonstrated a polymer based 2-domain YFDC modulator with Δβ-reversal at low modulation frequencies as a proof of concept [as shown in Fig. 1 (a), using a lumped element electrode], where we achieved an SFDR of $119dB/Hz^{2/3}$ with 11dB enhancement over the conventional MZI modulator [37]. Here, we present a traveling wave 2-domain directional coupler modulator to extend the high linearity to higher operational frequencies.

### A. Design

Fig. 1 (b) shows the schematic top view of our traveling wave 2-domain directional coupler modulator. A 1×2 multimode interference (MMI) 3-dB coupler (176.6μm×15μm) is designed to equally split the input optical power into the two arms of a directional coupler with total power transmission efficiency of 94%. Δβ-reversal can be easily achieved by domain-inversion poling on EO polymers. The directional coupler is divided into two domains, where the EO polymer in the first domain is poled in the opposite direction with respect to that in the second domain. A push-pull configuration is also applied, in which the two arms of the directional coupler in each domain are poled in opposite directions, to double the EO effect. Finally, the modulation electric field applied by a traveling wave electrode creates Δβ-reversal, which is indicated by red solid lines and green dashed lines in Fig. 1 (b).

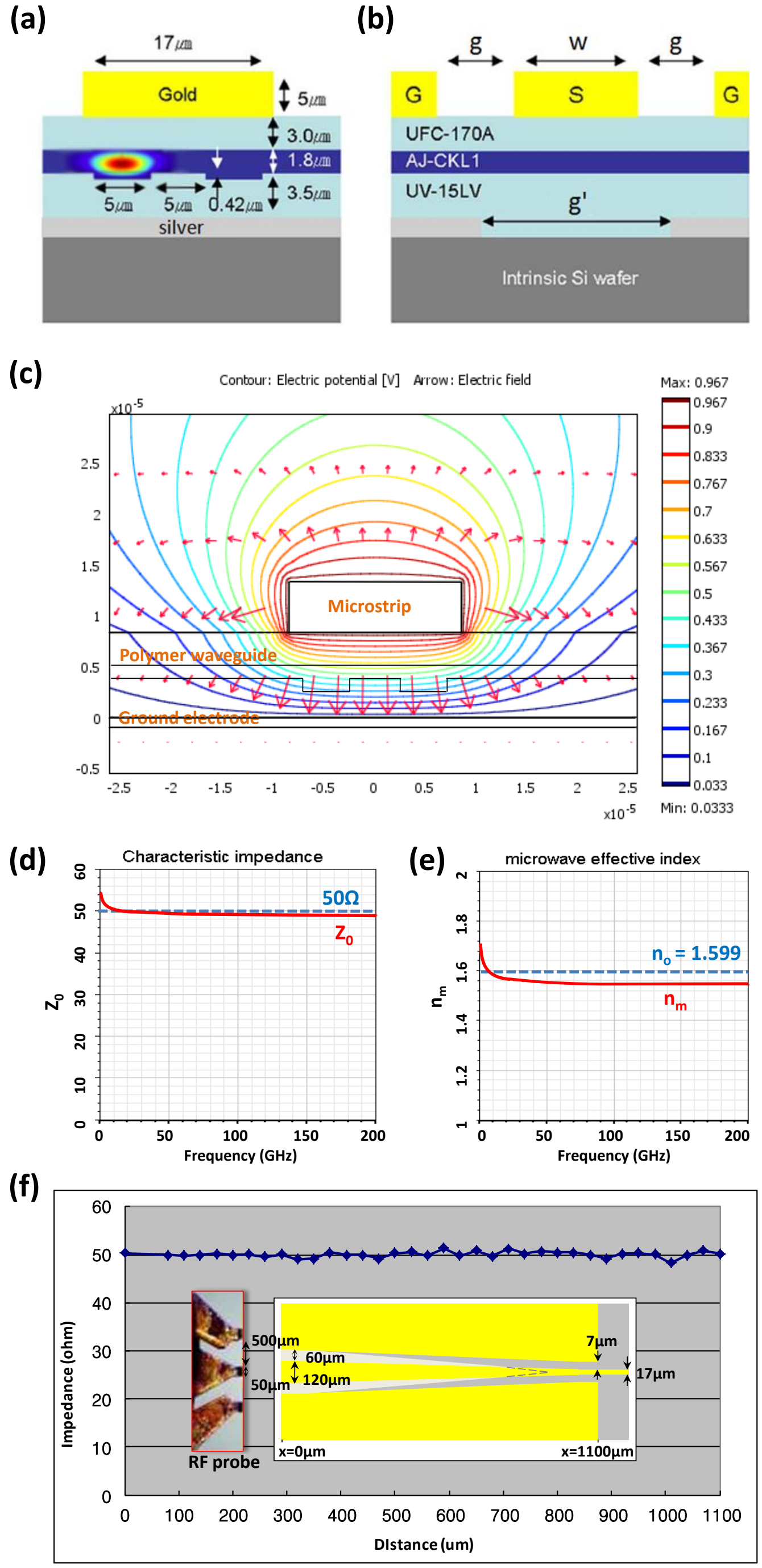

Fig. 2. (a) Cross section corresponding to A-A' of the traveling wave 2-domain directional coupler modulator in Fig. 1 (b), overlaid with optical mode profile in one arm. (b) Cross section corresponding to B-B' in Fig. 1 (b). S: signal electrode, G: ground electrode. (c) The schematic cross section of a microstrip line with design parameters overlaid with the contour of the normalized electric potential. The red arrows indicate the direction of electric field. (d) The characteristic impedance of the microstrip line over the frequency range 1-200GHz. The solid red curve indicates the characteristic impedance and the dashed blue line indicates 50Ω. (e) The microwave effective index of the microstrip line over the frequency range 1-200GHz. The solid red curve indicates the microwave effective index and the dashed blue line indicates the optical effective index of 1.599. (f) The top view of the quasi-CPW taper, matching the size of a microprobe. The characteristic impedance (at 10GHz) is matched with 50Ω along the microstrip-to-CPW transition direction.

The IMD3 suppression of a directional coupler modulator is a sensitive function of the normalized interaction length ($S_i$), defined as the ratio of the interaction length ($L_i$) of the $i^{th}$ section to the coupling length ($L_c$). Relative IMD3 suppression of a 2-domain directional coupler modulator can be graphically represented by plotting the calculated IMD3 suppression on ($S_1$,$S_2$) plane [39]. The condition $S_1$=$S_2$=2.86 provides excellent linearity as well as very high modulation depth [38], and is chosen for demonstration in this work. For our device, the total interaction length ($L_1$+$L_2$) of the directional coupler and the coupling length for the fundamental transverse magnetic (TM) mode are 2cm and 3.496mm, respectively. The coupling length is matched by tuning the design parameters of the trench waveguides, such as the core thickness and trench depth, using FIMMWAVE. Fig. 2 (a-b) shows the final cross sectional dimensions of the optical waveguides consisting of three layers of fluorinated polymers (bottom cladding: UV15LV, n=1.50; core: AJCKL1/APC from Soluxra, $\gamma_{33}$=80pm/V, n=1.63; top cladding: UFC170A, n=1.49). In addition, silver is selected as the ground electrode material since its smooth surface helps reduce the waveguide sidewall roughness originating from the scattering of ultraviolet (UV) light in photolithography, and its low resistivity is also beneficial to suppress the microwave conductor loss.

To extend the highly linear modulation to GHz frequency regime, a high-speed traveling wave electrode is designed. Some basic requirements for the design of traveling wave electrode include 1) impedance matching between the microwave guides and external electrical connectors, 2) velocity matching between the microwaves and optical waves, 3) low electrical loss in the microwave guides, and 4) electric field matching [40, 41] in the coupling between different transmission line structures.

In our device structure, considering the alignment of the RF modulation field with the direction of the $\gamma_{33}$ in the poled EO polymer film, which is in vertical direction, a microstrip line is a natural choice for the best RF-optical signal overlap integral. Fig. 2 (c) shows the schematic cross section of the designed gold microstrip line overlaid with the contour of normalized electric potential calculated by using COMSOL Multiphysics. It can be seen that both the arms of the directional coupler are under the effect of a uniform modulation field between the microstrip line and the ground electrode, and hence, the overlap integral between the optical mode and the RF modulation field is maximized. The frequency-dependent characteristic impedance and microwave effective index of the microstrip line can be numerically calculated by using ANSYS HFSS to match 50Ω and optical effective index of 1.599, respectively. Conductor loss and dielectric loss are considered in the calculation so that the results are accurate enough and close to the real case. Given the relative dielectric constant, $\varepsilon_r$=3.2, the gap between top and ground electrodes, h=8.3μm, and the microstrip thickness, t=5μm, the characteristic impedance of 50Ω can be matched when the microstrip width is w=17μm. As shown in Figs. 5 (d-e), over the frequency range of 1-200GHz, the characteristic impedance varies within 49-54.5Ω and the microwave effective index varies within 1.54-1.7. The bandwidth-length product due to the velocity mismatch is calculated as [42-45]

$$f \cdot L \cong \frac{1.9c}{\pi\left|n_m - n_o\right|}, \qquad (1)$$

where, f is the modulation frequency, L is the interaction length, c is the speed of light in vacuum, $n_m$ is the microwave effective index of the microstrip line, and $n_o$ is the optical effective refractive index of polymer waveguide. Using (1), we theoretically calculate the bandwidth-length product to be 306GHz·cm, corresponding to a modulation frequency limit of 153GHz for a 2cm-long microstrip line.

In addition, to couple the RF power from a ground-signal-ground (GSG) microprobe (e.g. Cascade Microtech with probe tip width of 50μm) into the 17μm-wide microstrip line with minimum coupling loss, a 1.1mm-long quasi-coplanar waveguide (CPW) taper is designed, as shown in Fig. 2 (f). The width and the gap of the coplanar waveguide [w and g in Fig. 2 (b)] are gradually changed along the taper to match the dimensions of the RF microprobe. Unlike the conventional CPW, the ground electrode under the taper is partially removed, and the bottom gap [see g´in Fig. 2 (b)] is gradually tuned along the taper based on the ground shaping technique [40, 41], so that there is a smooth transformation of the electric field profile in the CPW-to-microstrip transition to minimize coupling loss whereas the 50Ω is matched at all points along the transition direction, as shown in Fig. 2 (f).

### B. Fabrication

The device is fabricated on an ultra-high resistivity silicon wafer. A 1μm-thick silver film is patterned as the ground electrode by a lift-off process. A polymer trench waveguide is fabricated by spincoating, photolithography and reactive iron etching (RIE), in which the EO polymer is formulated by doping 25wt% of AJCKL1chromophore into amorphous polycarbonate (APC). 150nm-thick gold top electrodes are patterned by a lift-off process. 300nm-thick silicon dioxide is deposited on the entire surface of the device by an e-beam evaporation process to serve as a protection layer, so that push-pull poling can be done on EO polymer with poling electric field as high as 150V/μm at the glass transition temperature ($T_g$=145℃) without dielectric breakdown. After poling is finished, the poling electrodes are removed and then a 5nm-thick gold traveling wave electrode is fabricated by a constant-current electroplating process. The coplanar and ground electrodes are then connected with silver epoxy through via-holes. Finally, the device is diced and the waveguide facets are polished. Details about the fabrication process can be found in [46].

### C. Characterization

The performance of the fabricated traveling wave electrode is characterized by a vector network analyzer. Two air coplanar probes are used to couple RF power into and out of the tapered quasi-coplanar waveguides. The measured microwave loss of the traveling wave electrode over the frequency range 1-26GHz is shown in Fig. 3 (a). For reference, the theoretical electrode loss calculated by using ANSYS HFSS is shown in Fig. 3 (b). It can be seen that the measured transmission loss is proportional to the square root of frequency, implying that the microwave

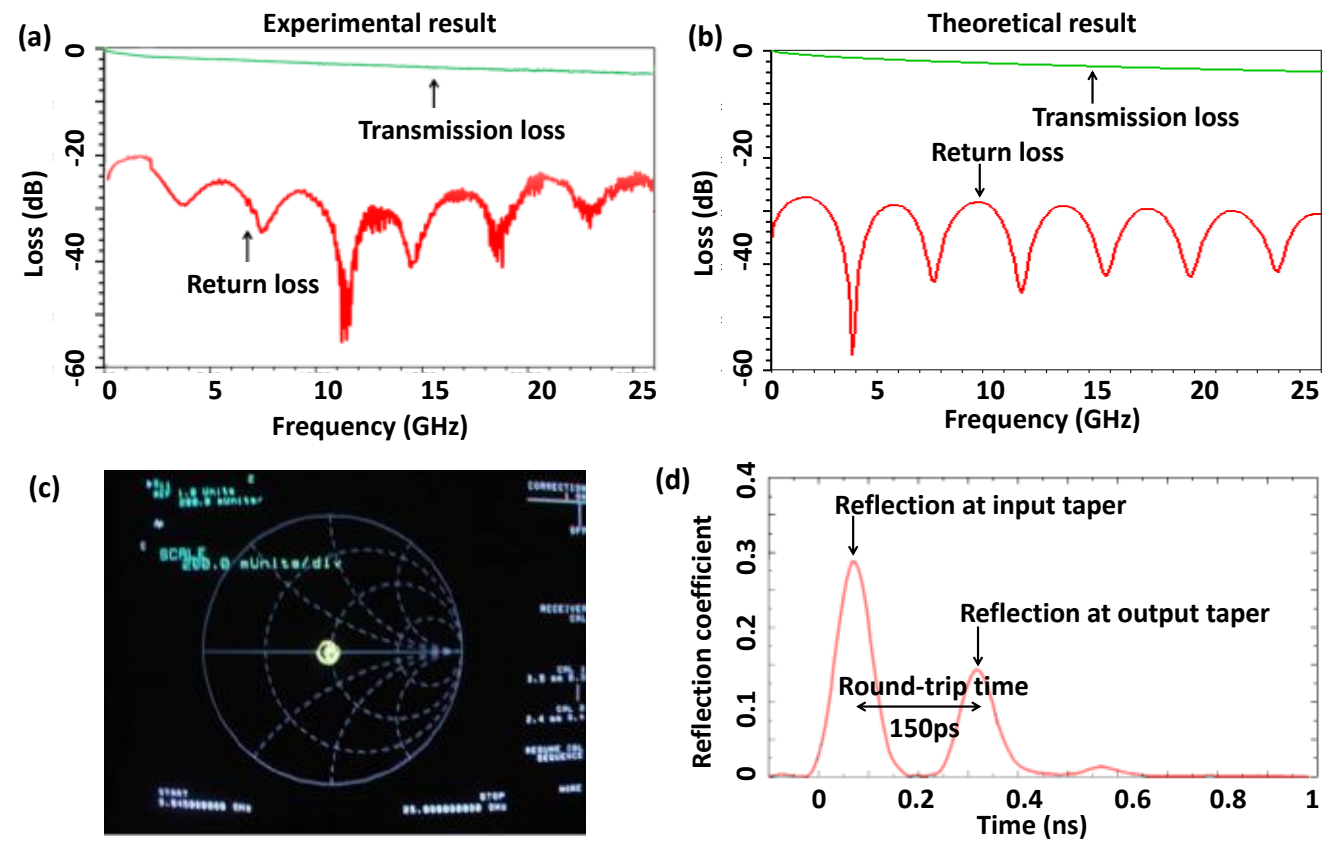

Fig. 3. Electrode characterization. (a) The measured transmission loss and return loss of the fabricated traveling wave electrode over the frequency range 1-26GHz. (b) The simulated transmission loss and return loss of the designed traveling wave electrode over the frequency range 1-26GHz. (c) The measured characteristic impedance of the fabricated traveling wave electrode is well centered at 50Ω on Smith Chart, indicating impedance matching. (d) The time domain measurement of reflection loss, for velocity matching demonstration.

Loss is dominated by the conductor loss (skin effect loss) of the electrode [47, 48] which is measured to be 0.65±0.05dB/cm/GHz$^{1/2}$. The 3-dB electrical bandwidth measured from transmission loss curve is 10GHz, nearly the same value as that from the theoretical calculation. This bandwidth is limited by the relatively low conductivity of the poorly electroplated gold electrode and can be enhanced by improving the electroplating quality. The measured return loss is well below -20dB. This low return loss is mainly due to the excellent impedance matching as well as the smooth electric field transformation in the CPW-microstrip-CPW transition section. It can be noticed that this value is still higher than the theoretical result (<-27dB), probably due to the fabrication errors.

It is shown in Fig. 3 (c) that the characteristic impedance is well centered at 50Ω on Smith chart, indicating impedance matching. The velocity matching between microwaves and optical waves is evaluated by the time domain measurement of the return loss, as shown in Fig. 3 (d). The effective relative dielectric constant of the microstrip line is measured to be 2.76 and the resulting index mismatch between microwave sand optical waves is 0.06. Then, the bandwidth-length product due to this velocity mismatch can be calculated by (1) to be 302GHz·cm, so the modulation frequency limit corresponding to 2cm interaction length would be 151GHz, which matches the theoretical calculation result [153GHz calculated from Fig. 2 (e)] pretty well.

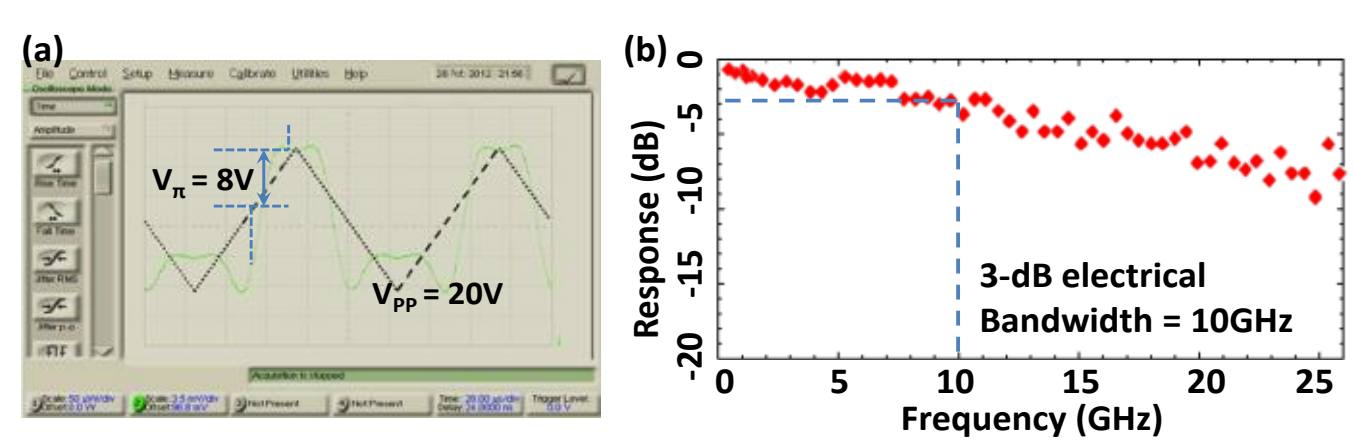

Fig. 4. Modulation measurement. (a) The transfer function of over-modulation with $V_{pp}$=20V at 10KHz (optical wavelength=1550nm). The half-wave switching voltage is measured to be $V_\pi$=8V. (b) The frequency response of the small signal modulation measured at 4% modulation depth. The 3-dB bandwidth is measured to be 10GHz.

In the modulation test, TM-polarized light with 1550nm wavelength from a tunable laser is butt-coupled into the waveguide through a single mode polarization maintaining fiber. The measured total optical insertion loss is 16dB, which includes propagation loss of 9dB (absorption loss of 2dB/cm for AJCKL1/APC and scattering loss of 1dB/cm over the total device length of 3cm), coupling loss of 6dB (3dB/fact times 2 facets), and 1 dB loss from the MMI splitter. This relatively high loss is attributed to the roughness of the 3cm long waveguide sidewalls generated in the RIE process and the roughness of input and output waveguide facets. To measure the $V_\pi$, an RF signal with $V_{pp}$=20V at 10KHz is used. The measured transfer function of over-modulation is shown in Fig. 4 (a). By finding the difference between the applied voltage at which the optical output is at a maximum and the voltage at which the optical output is at the next minimum, the $V_\pi$ is measured to be 8V at 10KHz, which is somewhat higher than expected probably due to the low poling efficiency of EO polymer and electrode loss. The frequency response of the device is evaluated by a small signal optical modulation measured at 4% modulation depth. RF signal from the network analyzer is fed into the traveling wave electrode through a GSG microprobe. The modulated optical signal is boosted by an erbium doped fiber amplifier (EDFA), converted to electrical signal by a photodiode, and then measured by a microwave spectrum analyzer. The frequency response measured at 4% modulation depth is presented in Fig. 4 (b), from which the 3-dB bandwidth of the device can be found to be 10GHz. This bandwidth is mainly limited by the conductor loss of the traveling wave electrode as mentioned before.

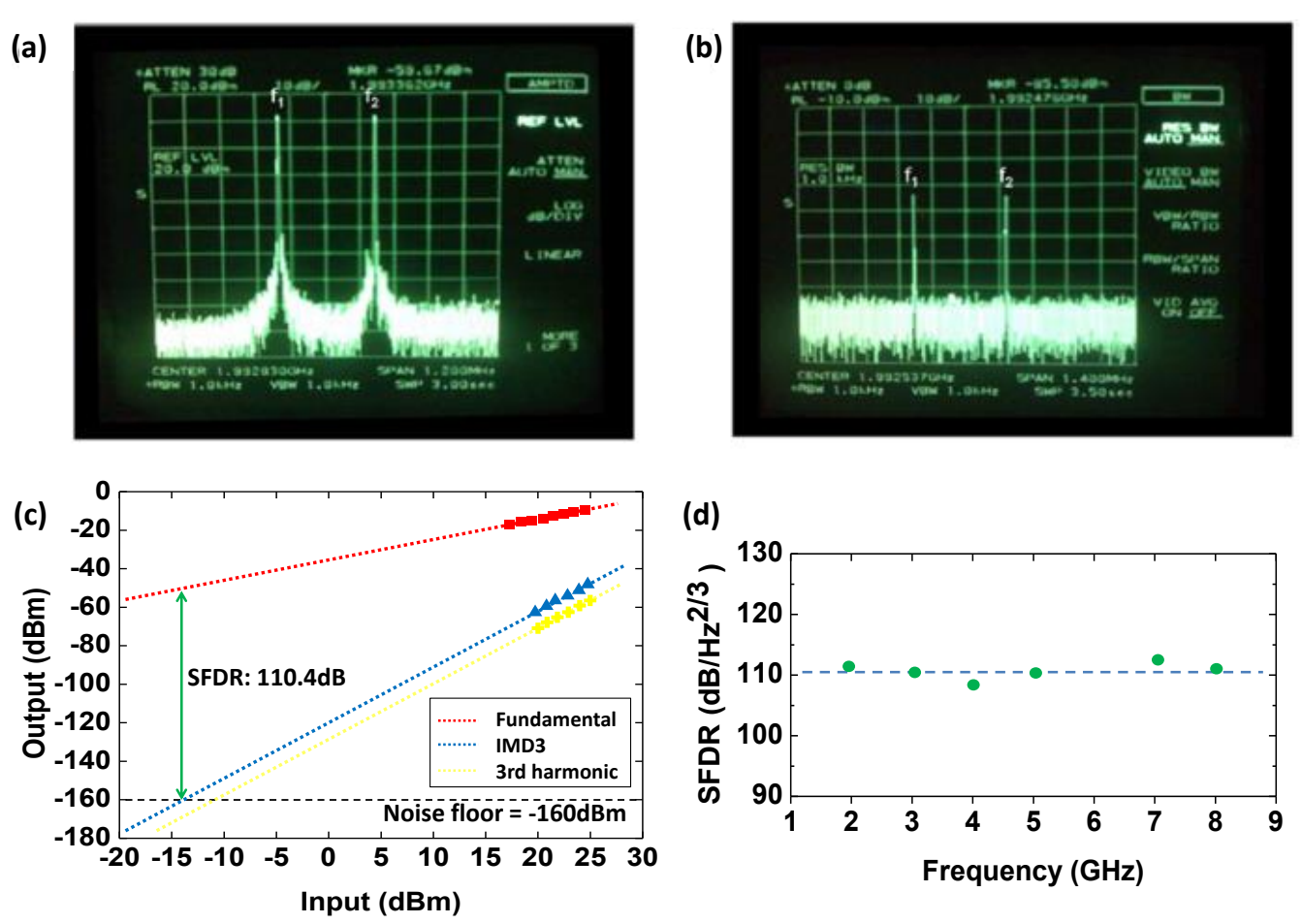

Fig. 5. Linearity evaluation. (a) Input two-tone signals ($f_1$and $f_2$) centered at 1.9928 GHz with 330 KHz tone-interval. (b) Measured output fundamental signals. (c) The response plot of fundamental signals, third-order intermodulation distortion signals, and third harmonic distortion signals measured at 8GHz. (d) Spurious free dynamic range measured at 2-8GHz.

A two-tone test is performed to evaluate the linearity of the device. A sweep oscillator is used as the second RF source for the two-tone input signals. Pre- and post-RF amplifiers are used. The two-tone input signals and the resulting output signals are shown in Fig. 5 (a-b), respectively. IMD3 signals, which are supposed to appear at one tone-interval away from the fundamental signals if present, are not observed in Fig. 5 (b). A possible reason is that the IMD3 signals are well suppressed and buried under the noise floor. The power level of the two-tone input signals is 12dBm as shown in Fig. 5 (a), which is the maximum level available in our two-tone test setup, and this power level translates into a modulation depth of 15%. Since the IMD3 suppression of the fabricated device is out of the measurable range in our two-tone test setup, SFDR is evaluated through an indirect method. A mono-tone test is done with the same modulation depth and under the same conditions as the two-tone test. It is found that the third harmonic distortion of our device comes in the detectable range at the mono-tone input signal level above 20dBm. Then, the IMD3 signals are obtained by adding 9.54dB to the measured third harmonic distortion signals [49]. The SFDR is measured by extrapolating the IMD3 plot to find an intercept point with the noise floor and then measuring the difference with the extrapolated fundamental signal as illustrated in Fig. 5 (c).

Considering the relative intensity noise of the distributed feedback laser and the shot noise of the photodiode, it is very difficult to achieve a noise floor below -145dBm in real analog optical links [38]. However, laboratory test results in most literatures are frequently presented assuming the noise floor at -160dBm considering the typical fiber-optic link parameters [50-52]. By using -160dBm as noise floor, our measured SFDR is within 110±3dB/Hz$^{2/3}$ over the modulation frequency range 2-8GHz, as shown in Fig. 5 (d). The low end frequency is determined by the operation range (2-26.5GHz) of the pre-amplifier and the high end is limited to 8GHz because the third harmonic of the modulation frequency above 8GHz goes beyond the scope (~26.5GHz) of the microwave spectrum analyzer. The SFDR at 6GHz is missing due to the irregular gain of the post-amplifier at 18GHz. As a comparison, Schaffner et al. reported the SFDR of 109.6dB/Hz$^{2/3}$ at 1GHz with a $LiNbO_3$ directional coupler modulator which is linearized by adding passive bias sections [32]. In their measurement, the noise floor was set at -171dBm, which offers 7.3dB extra dynamic range compared with the noise floor at -160dBm. Hung et al. achieved even higher SFDR of 115.5dB/Hz$^{2/3}$ at 3GHz with a linearized polymeric directional coupler modulator by subtracting the distortions of the measurement system [52]. Dingel et al. demonstrated the SFDR of 130.2dB/Hz$^{2/3}$ by using a unique combination of phase modulator and a weak ring resonator modulator within a MZI structure [53]. In this measurement the specific modulation frequency is not clearly mentioned, although it is stated that the broadest possible bandwidth can be 20GHz. Here, our SFDR of 110±3dB/Hz$^{2/3}$ includes the distortions from the entire measurement system as well as the device. To the best of our knowledge, such high linearity is first measured at a frequency up to 8GHz by our group.

## III. Silicon/Polymer Hybrid Photonic Crystal Waveguide Modulator with Low Switching Voltage

Hybrid silicon and EO polymer photonic devices can benefit from the large EO coefficient of the polymer as well as the compact size made possible by the large index of silicon [54-57]. For example, a silicon-organic hybrid electro-optic modulator was demonstrated to operate at 42.7 Gbit/s, by applying a novel electron accumulation-layer technique to increase the conductivity of the thin silicon electrodes [58]. A transmission line driven slot waveguide MZI modulator was demonstrated to achieve a record low driving voltage of $V_\pi$<200mV at 10GHz [59]. Utilizing the slow light effect, photonic crystal waveguides (PCWs) refilled with EO polymers can further reduce the device size [60]. Here, we present the design and experimental demonstration of a MZI based on EO polymer refilled silicon slot PCW modulators.

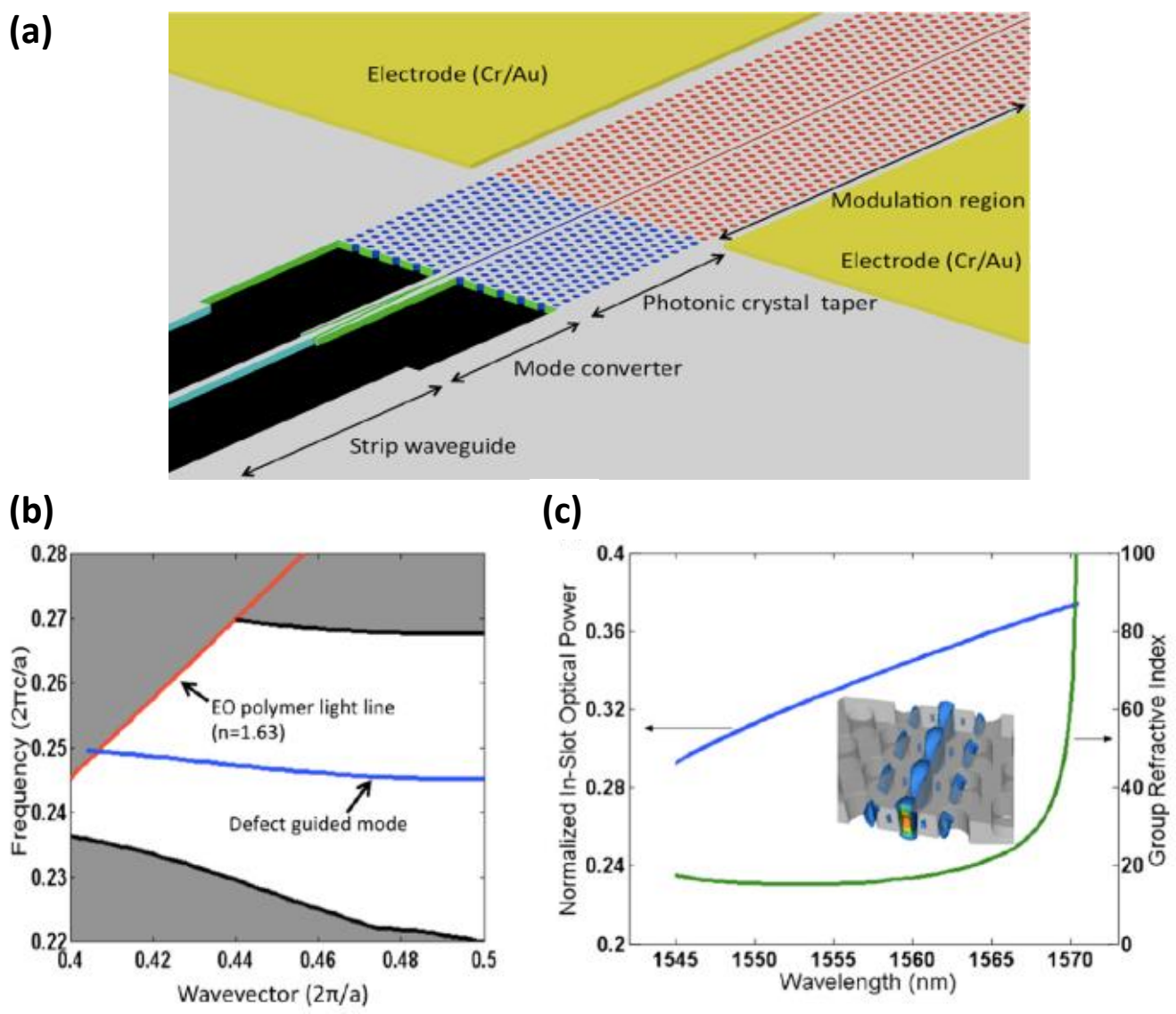

Fig. 6. (a) Schematic of the input strip waveguide, optical mode converter, PCW taper, and modulation region. (b) Enlarged portion of the dispersion diagram for the guided mode. (c) Group index and normalized in-slot optical power of the guided mode as a function of the optical wavelength. Optical mode profile at $n_g$=100 is shown in inset.

## A. *Design*

Fig. 6 (a) shows a schematic of the slot PCW modulator. Input and output waveguides are conventional silicon strip waveguides connected to a slot PCW through optical mode converters [61]. The slot PCW is formed by replacing one row of the holes with a narrow slot width, $S_w$=75nm. Compared to other devices with slot width over 120nm [30, 62], the narrower slot provides higher modulation efficiency at the same driving voltage, as well as good optical confinement, without compromising the EO polymer infiltration.

The modulation region, with slot nanostructures, is formed in a hexagonal lattice slab PCW with lattice constant a=385nm and hole diameter d=217nm, which has a total length of 308 μm. Silicon slot PCW region, including the slot and air holes, is fully covered by EO polymer, AJCKL1/APC ( $\gamma_{33}$=80pm/V, n=1.63) [63]. The dispersion diagram of the fundamental defect-guided mode is shown in Fig. 6 (b), which is calculated by RSoft BandSOLVE. The group index, $n_g$, of this guided mode as a function of wavelength is shown in Fig. 6 (c), which shows that $n_g$ can exceed 100 when the wavelength is tuned close to the band edge of 1569nm. The optical intensity profile ($|E|^2$) of the guided mode at $n_g$=100 is shown in the inset of Fig. 6 (c). Fig. 6 (c) also shows the fraction of total guided mode power in the slot, Γ=0.37, calculated over one complete period of the fundamental guided mode profile in the slot PCW by RSoft BandSolve simulation. It should be noted that the integration of 75nm slot into the PCW causes light with high $n_g$ within the defect mode spectrum to remain concentrated in the slot, which will otherwise penetrate to second or third row of holes [64]. Additionally, the silicon is slightly doped to function as an electrode. Compared to the case with undoped silicon, the electric field across the slot can be increased, thus enabling better RF interaction within the EO polymer infiltrated in the slot. To effectively couple light into the slow light region, a PCW taper from W1.08 to W1.0 (the air hole spacing outside the slot is changed from dW=1.08√3a to dW=√3a) is designed to minimize the group index mismatch between the strip waveguide and the slot PCW [65, 66], as shown in Fig. 6 (a).

## B. *Fabrication*

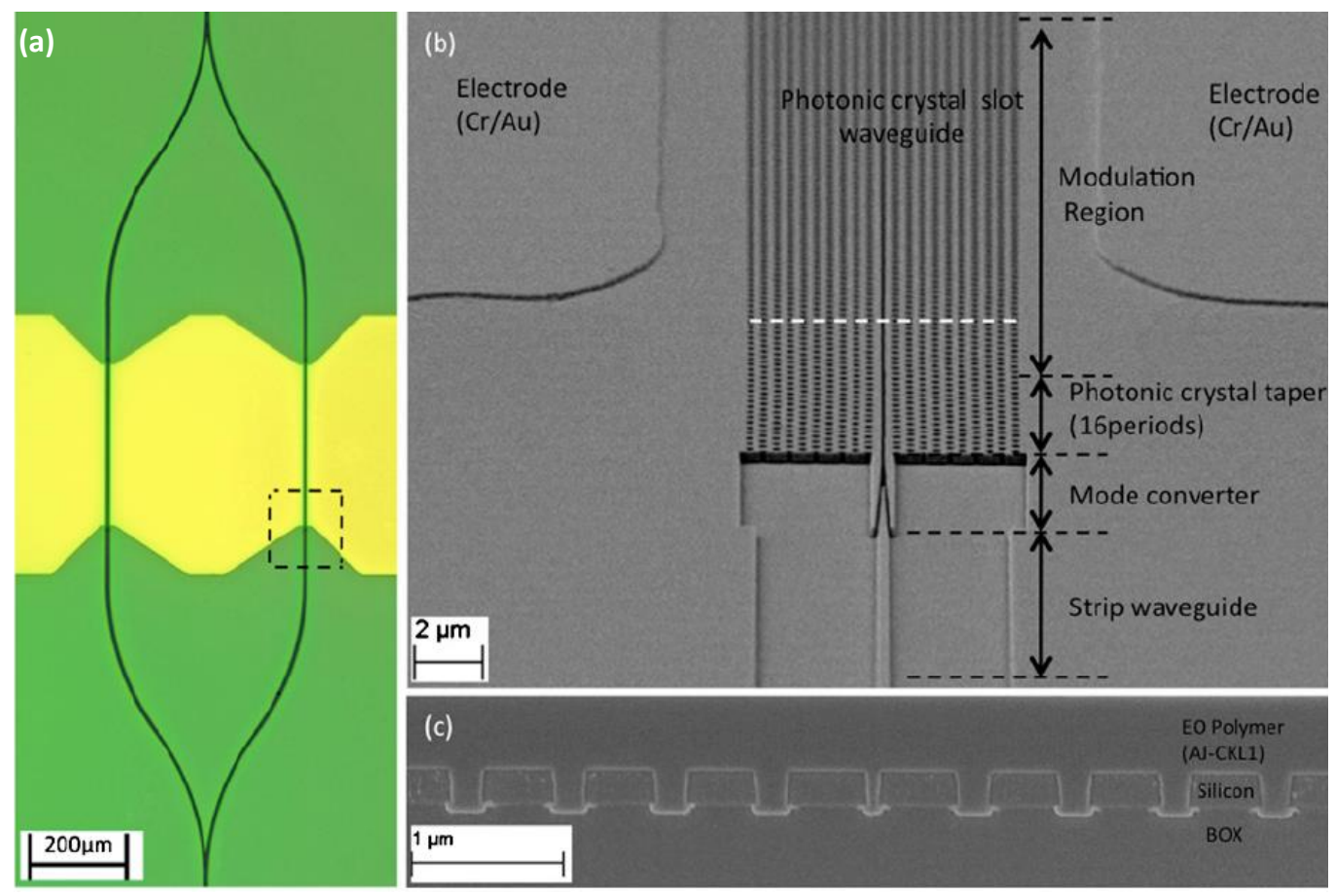

Fig. 7. (a) Optical microscope picture of the fabricated MZI structure. (b) SEM picture showing the enlarged view of the dotted square area in (a). (c) Cross-sectional SEM picture taken across the dotted line in (b) after covering the entire structure in (a) with EO polymer. Complete infiltration of EO polymer into the 217nm air holes and 75nm slot is confirmed.

This hybrid nanophotonic modulator is fabricated on a SOI wafer with 230nm slightly doped top silicon and 3 μm buried oxide. Details of fabrication are described in [65]. The device is fabricated by using electron-beam lithography and RIE in a single patterning/etch step, whereas the gold electrodes are patterned by photolithography and lift-off processes [Fig. 7 (a)]. Fig. 7 (b) shows the scanning electronic microscopy (SEM) image of the silicon slot PCW MZI modulator. The EO polymer is infiltrated into the slot PCW waveguides by spincoating. Fig. 7 (c) shows cross sectional view of the 75nm slot PCW completely refilled with EO polymer.

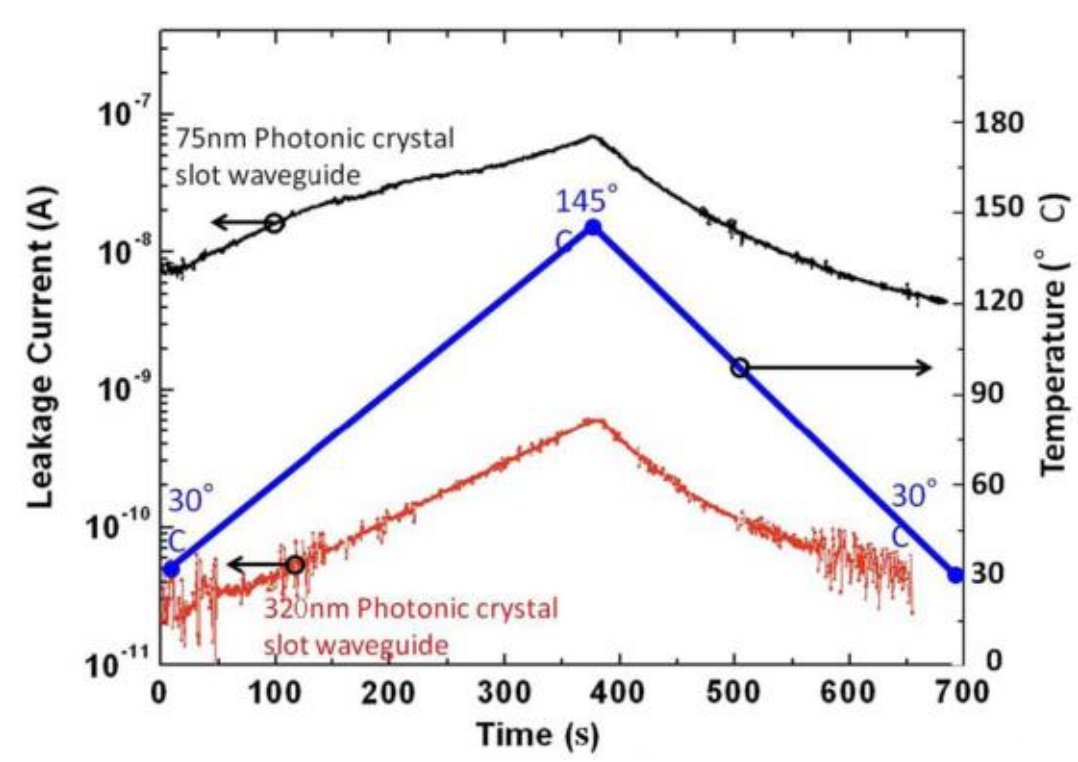

Fig. 8. Temperature-dependent leakage current during the process of poling on EO polymer (AJCKL1/APC) refilled slot PCWs with slot width of 75nm (black) and the 320nm (red), respectively. The blue curve indicates the change of temperature.

After EO polymer infiltration, the sample is heated to the $T_g$=145 ℃ while a 200V/μm poling field is applied. Upon reaching the glass transition temperature, the sample is then cooled down to room temperature, and the poling voltage is switched off. The leakage current across this 75nm slot as well as the hot plate temperature during the poling process is monitored in situ and shown in Fig. 8.

The leakage current is known to be detrimental to the poling efficiency [67]. In order to reduce the leakage current, we investigate the effects of the slot PCW structure on the poling efficiency. Especially, we design and fabricate a EO polymer refilled PCW with a 320nm-wide slot, in which the leakage current is reduced by over 2 orders of magnitude compared to that in the narrow slot (75nm wide) as shown in Fig. 8. We use this finding for designing a more efficient structure in Section IV.

### C. *Characterization*

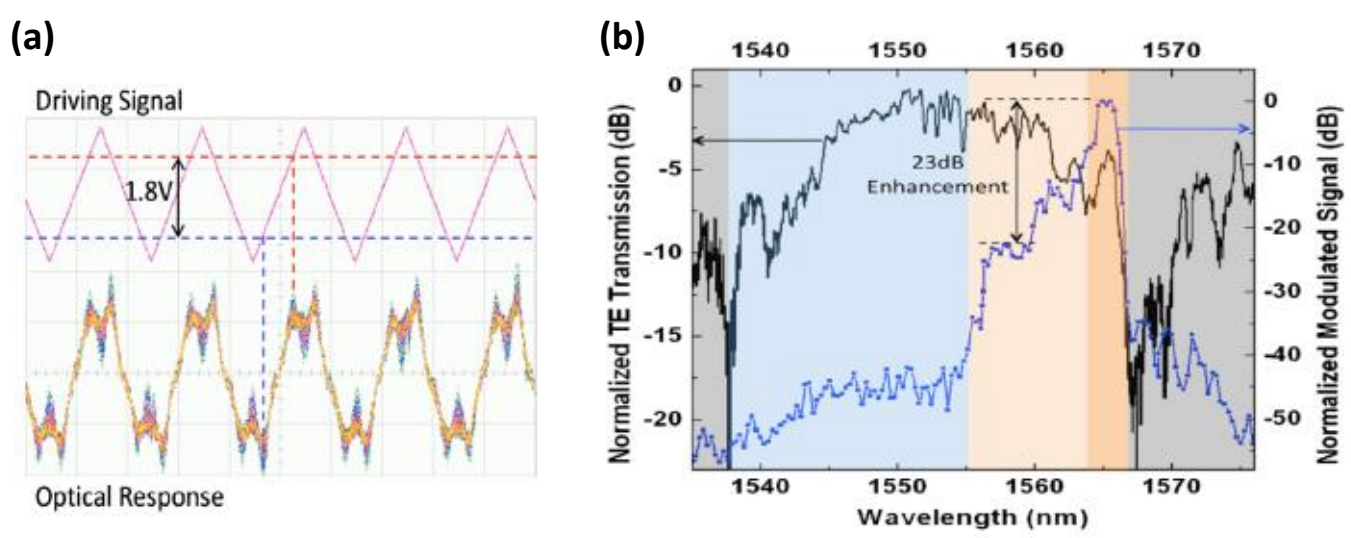

Fig. 9. (a) Modulation transfer-function measurement at 50KHz (optical wavelength=1564.5nm). Upper: applied voltage; lower: optical output signal. The half-wave switching voltage is measured to be $V_\pi$ =1.8V. (b) The wavelength dependence of the normalized modulated optical signal (blue) and normalized optical transmission (black). Four distinct regions are shown in this figure: 1) normal group velocity region with high optical transmission and low modulated signal (blue); 2) transitional region with gradually decreasing optical transmission and rapidly increasing modulated signal (light orange); 3) slow light region with relatively low optical transmission but extremely high EO modulation (orange); 4) photonic band gap and beyond with minimized modulation (gray).

To characterize the modulator performance, transverse electric (TE) light from a broadband amplified spontaneous emission (ASE) source is butt coupled into the modulator with a polarization maintaining tapered lensed fiber. The transmitted light is collected by a single mode lensed fiber and analyzed with an optical spectrum analyzer. We observed a 5nm deviation in the photonic band edge at 1569nm compared to simulation results, which is attributed to fabrication errors. A laser source is tuned to 1564.5nm, corresponding to the slow light region, where a maximum modulation response is achieved. The modulator is biased at the 3dB point and driven by a 50KHz triangular wave. The modulated optical signal is converted to electrical signal by a gain-switchable photodetector. Fig. 9 (a) shows that the modulator has a $V_\pi$ of 1.8V. The effective EO coefficient can be calculated as follows,

$$\gamma_{33,effective} = \frac{\lambda S_w}{n^3 V_\pi \Gamma L} \;, \qquad (2)$$

where, wavelength λ=1565nm, slot width $S_w$=75nm, EO polymer index n=1.63, interaction length L=308 μm, $V_\pi$=1.8V, and Γ=0.37, so the effective in-device $\gamma_{33}$=132pm/V. The device also achieves very high modulation efficiency $V_\pi \times L$=1.8V ×308 μm=0.56V mm. This result is nearly one order of magnitude lower than that reported in [68].

To confirm the dramatic EO modulation enhancement due to slow light effect, all testing conditions are fixed and the wavelength is tuned from 1535 to 1575 nm. The wavelength dependence of normalized modulated signal under 1 V of driving voltage is plotted in Fig. 9 (b), together with normalized optical transmission spectrum of the EO polymer refilled slot PCW modulator. The defect-guided mode of slot PCW occurs from 1538 to 1567 nm. Although the normalized optical transmission reaches maximum at 1550 nm, the normalized modulated signal is only about 45dB. As we tune to longer wavelengths, the intensity of the modulated signal increases dramatically due to slow light enhancement. The peak modulated signal around 1565 nm is 23dB higher than in the transitional region, where the photodetector starts to measure sensible modulation response. Above 1566 nm, the modulated signal decreases sharply due to transmission cut-off by the photonic band gap.

## IV. Introducing Band-Engineering into Silicon/Polymer Hybrid Photonic Crystal Waveguide Modulator for Small Wavelength Dispersion

### A. *Design*

In this section, the design principle is based on the fact that a wider slot waveguide will significantly suppress the leakage current and thus can lead to higher poling efficiency. The basic silicon slot PCW structure is schematically similar to the one shown in Fig. 6 (a), but with a thickness of t=250nm, a lattice constant of a=425nm, a hole diameter of d=300nm, and a total photonic crystal length of 314μm (including 300μm-long active length and two 7μm-long PCW tapers). A W1.3 waveguide (the air hole spacing outside the slot is dW=1.3√3a) is chosen. A new EO polymer material, SEO125/APC from Soluxra, LLC, consisting of a guest/host system of 25%weight chromophore SEO125 into amorphous polycarbonate (APC) is used in this device. This EO polymer has the same refractive index (n=1.63) as AJCKL1 used in Section III, but has higher EO coefficient, $\gamma_{33}$=125pm/V. The optical intensity profiles ($|E|^2$) of the guided mode at the band edge (wave vector of π/a) is numerically calculated to have in-slot power fraction around Γ=0.33.

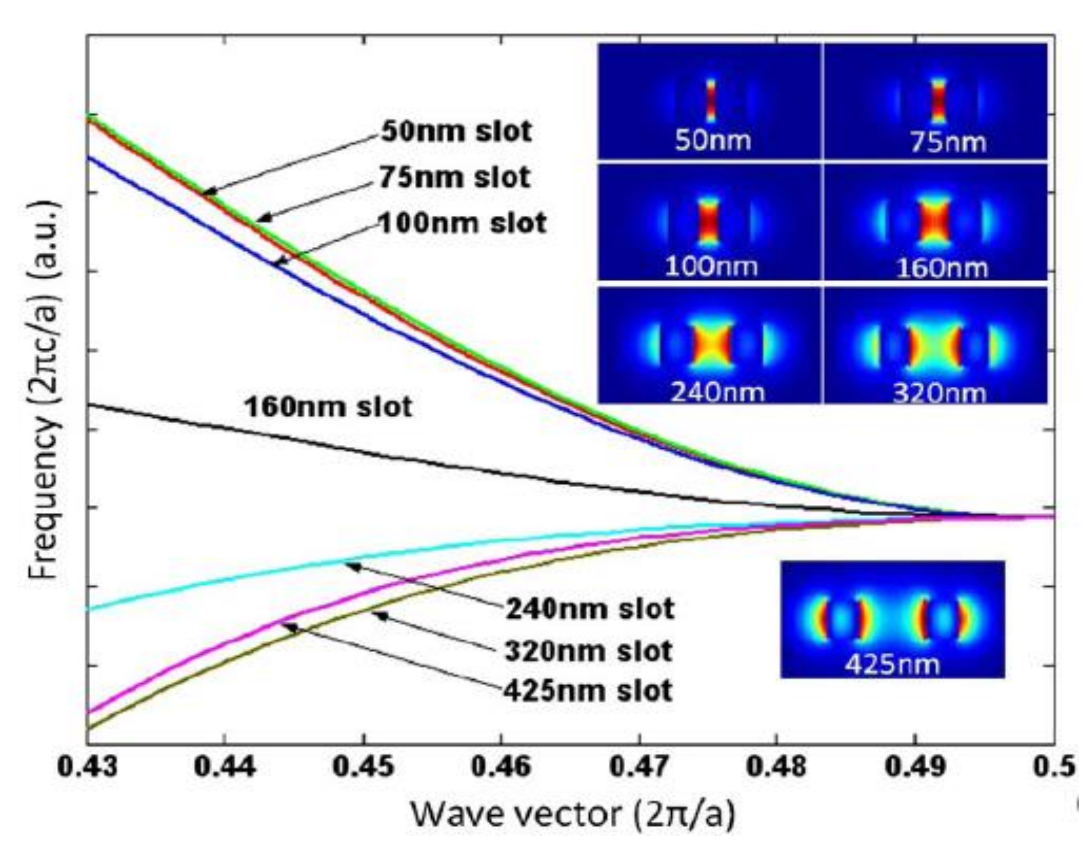

Fig. 10. Photonic band diagrams and optical mode profiles of EO polymer refilled slot PCWs with different slot widths.

The effect of slot width in silicon slot PCW is investigated. Fig. 10 shows the simulated band diagrams of slot PCW with different slot widths in conjunction with two-dimensional cross-sectional views of the optical intensity profiles of the guided modes at the photonic band edges. As the slot width increases the slot mode is no longer supported, because the slot mode inclines to be decoupled into two separate waveguide modes, as shown in Fig. 10. Additionally, large slot width is detrimental to EO modulation efficiency because the electrode separation is increased. Therefore, we believe that a 300-350nm slot width with 30% in-slot power is an optimized design for an EO polymer refilled silicon slot PCW modulator. Here, we choose slot width of 320nm for demonstration. The refractive index of the polymer is assumed to be independent from the slot width.

In addition, band-engineering is used to achieve low-dispersion slow-light (constant group velocity) propagation in this slot PCW for the MZI operation [69, 70]. Specifically, we have chosen the lateral lattice shifting approach [71] for the following advantages. First, all the PCW holes are the same and this increases the fabrication yield and reproducibility compared to techniques that require precise control of multiple hole diameters. Also, it facilitates targeting a desired group velocity over a bandwidth of interest since these two parameters can be tuned relatively independently compared to longitudinal lattice shifting. Finally, it does not change the defect line width and facilitates efficient coupling between the fast-light mode of a silicon slot waveguide (group index, $n_g$ ~3) and the slow-light mode in the slot PCW ($n_g$ >10).

A schematic of the band-engineered slot PCW is shown in Fig. 11 (a). The first three adjacent rows on each side of the defect line are shifted parallel to the line defect to modify the dispersion diagram of the defect mode. Using Rsoft BandSolve module, we simulate the fundamental guided defect mode profile and band structure of this band-engineered slot PCW, as shown in Fig. 11 (a-b). For lattice constant, a=425nm, it is found that with hole diameter, d=300nm, lattice shifting step $s_1$=0, $s_2$=-85nm, $s_3$=85nm, slot width of $S_w$=320nm, and dW=1.54(√3)a, we can achieve an average $n_g$ of 20.4 (±10%) over 8.2nm bandwidth as shown in Fig. 11 (c). Note that the second and third rows are shifted in different directions. The absolute value of the group velocity dispersion (GVD) remains below 10ps/nm/mm over the entire bandwidth, as shown in Fig. 11 (d).

In order to efficiently couple light into and out of the slot PCW, a PCW taper has been designed. The PCW taper consists of a mode converter, as shown in Fig. 6 (a), and a non-band-engineered PCW (a=425nm, d=300nm, $s_1$=0, $s_2$=0, $s_3$=0, $S_w$=320nm), for which, the width of the line defect (dW) parabolically increases from dW=1.45(√3)a to dW=1.54(√3)a. The slot width ($S_w$), hole diameter (d), and the period (a) remain constant. The band structure and the group index variation of the PCW taper are shown in Fig. 11 (b-c), which can be compared with those of the band-engineered slot PCW. An $n_g$ of ~6 of the PCW taper over the optical bandwidth of interest provides an effective interface for coupling between the slow light mode of the slot PCW ($n_g$=20.4) and the fast light in the silicon slot waveguides ($n_g$ ~3). In other words, the PCW taper gradually increases (slows down) the group index (propagating light) from the interface with the mode converter to the interface with the high $n_g$ slot PCW.

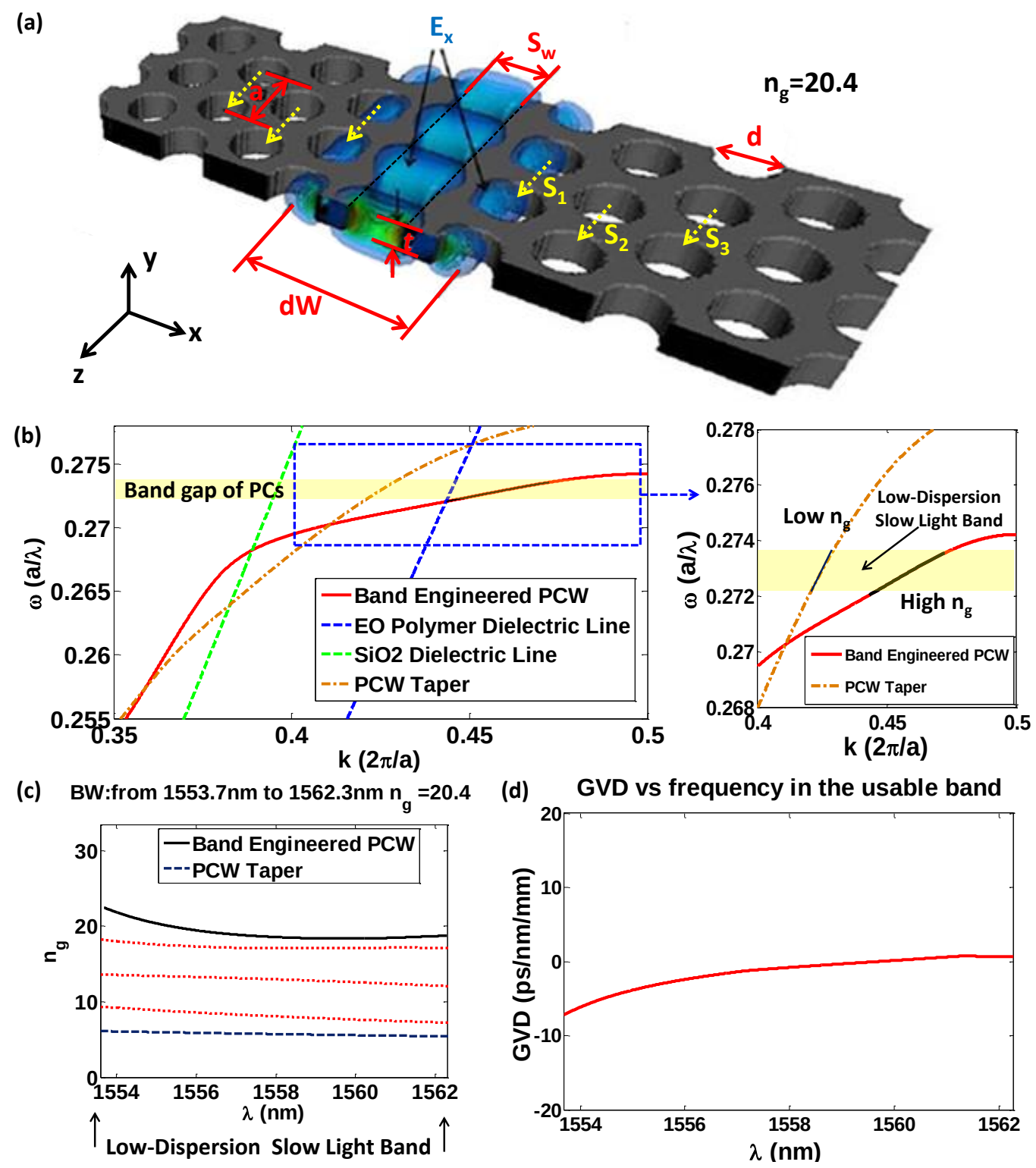

Fig. 11. (a) A 3D schematic of band-engineered slot PCW, overlaid with the 3D electric field profile of the fundamental guided defect mode. (b) The band structure for 2 PCWs, the band-engineered slot PCW (a=425nm, d=300nm, $s_1$=0, $s_2$=-85nm, $s_3$=85nm, $S_w$=320nm, dW=1.54(√3a), and the PCW taper (a=425nm, d=300nm, $s_1$=0, $s_2$=0, $s_3$=0, $S_w$=320nm, dW=1.45(√3a). The "flattening" of the mode in the band-engineered slot PCW can be noticed. The black curve highlights the low dispersion slow light section of the mode of the band-engineered slot PCW. The dielectric light lines corresponding to the $SiO_2$ (n=1.45) and EO polymer (n=1.63) cladding layers are shown. The useful part of the mode falls below the both light lines. (c) Variation of the group index vs. wavelength for the band-engineered slot PCW and the PCW taper. (d) Variation of the group velocity dispersion v.s. wavelength for the band-engineered slot PCW.

The required change of effective index of the EO polymer for the optical modulator to achieve π phase shift is given as $\Delta n = 1/(2\Gamma)\times(n/L)\times\lambda/n_g = 1/(2\times0.33)\times(1.63/300\,\mu m)\times1550nm/20.4 = 0.000625$. This change of EO polymer index can be realized by applying a half-wave switching voltage of $V_\pi = 2S_w\Delta n/(n^3\gamma_{33}) = 0.853V$, where $\gamma_{33}$=100pm/V is the EO coefficient of the polymer. Given the potentially large $\gamma_{33}$=125pm/V of EO polymer SEO125 and demonstrated high poling efficiency achievable in wide slots (320nm compared to conventional 100nm) [72], the estimated $\gamma_{33}$=100pm/V here is a realistic value. Therefore, from these calculations, the theoretical $V_\pi\times L$= 0.853V×300μm=0.256V mm. The expected effective in-device $\gamma_{33}$ is then calculated by (2) to be 1365pm/V, where λ=1560nm, $S_w$=320nm, n=1.63, L=300μm, Γ=0.33.

In addition, different from the slot PCW modulator in Section III, multimode interference (MMI) couplers are used for beam splitting and combining [73], a through-etched subwavelength grating (SWG) with over 50% efficiency are designed to couple light into and out of the silicon strip [74].

### B. Fabrication and characterization

The fabrication procedure for the modulator is the same as that described in Section III. The only difference is the EO polymer post-processing. The SEO125/APC is poled by a poling electric field of 100V/μm. After the device is rapidly heated up to the $T_g$=150 ℃, the temperature is held for 1min before cooling down.

To characterize the slot PCW, light from a broadband ASE source is coupled into the slot PCW via grating couplers. The optical output is measured by an optical spectrum analyzer (OSA). Fig. 12 (a) shows the measured transmission spectrum. A clear band gap with more than 30dB contrast is observed. The stepwise slope at the band edge indicates efficient coupling into the slow-light modes of the slot PCW which benefits from PCW taper. Otherwise, the slope would be more curved due to the larger index mismatch. The Fabry-Perot oscillations are measured to be smaller than 1dB over the low-dispersion wavelength range of ~7nm.

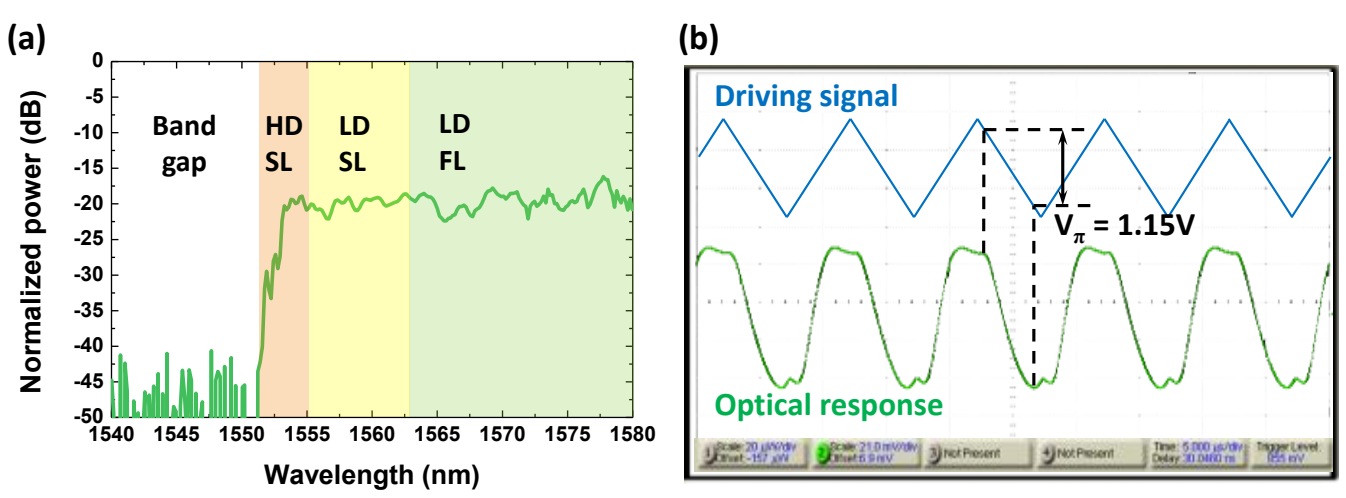

Fig. 12 (a) The transmission spectrum of the band-engineered slot PCW refilled with EO polymer. SL: slow light; FL: fast light; HD: high dispersion; LD: low dispersion. (b) Transfer function of over-modulation at 100KHz (optical wavelength =1558nm). The measured $V_\pi$=1.15V.

For the modulation test, a tunable laser is used. TE light from the laser source is tuned to a wavelength of 1558nm, where maximum modulation response is achieved. The total optical insertion loss is 20dB, including the 6.5dB/facet coupling loss from grating couplers. The modulator is biased at the 3dB point and driven by a 100KHz triangular RF wave with a peak-to-peak voltage of 1.2V. The output optical waveform measured by a digital oscilloscope in Fig. 12 (b) shows that over-modulation occurs at 1.15V, which is the $V_\pi$ of the modulator. The effective in-device $\gamma_{33}$ is then calculated by (2) to be 1012pm/V. This extraordinarily high $\gamma_{33}$ value confirms the combined enhancing effects of slow light and an improved poling efficiency. The effective in-device $\gamma_{33}$ remains over 1000pm/V over 5nm wavelength range. With the designed $n_g$=20.4, we estimate the in-device $\gamma_{33}$ to be 74pm/V, significantly more than the 59pm/V in our another work on a non-band-engineered slot PCW [72].

This band-engineered 320nm slot PCW modulator also achieves very high modulation efficiency with $V_\pi \times L$=1.15V×300μm=0.345V mm. This $V_\pi \times L$ shows 38% improvement over the $V_\pi \times L$=0.56V mm in the 75nm slot PCW modulator in Section III [31], and 22% improvement over the $V_\pi \times L$= 0.44V mm in our previous non-band-engineered 320nm slot PCW modulator [72].

In summary, we demonstrate a band-engineered EO polymer refilled slot PCW MZI modulator. The slow-light effect and the improved poling efficiency of high performance EO polymer makes possible effective in-device $\gamma_{33}$ of 1012pm/V and $V_\pi \times L$ of 0.345V mm. To the best of our knowledge, this is the best figure of merit that has ever been reported. Table I shows the comparison of our results with some other group's results in recent years [29-31, 62, 72].

Table I. Record of silicon/polymer hybrid EO modulators in recent years

| Year | Device Picture | Slot Width | Device Performance | E-O Polymers | Research Group |
|---|---|---|---|---|---|
| 2008 | 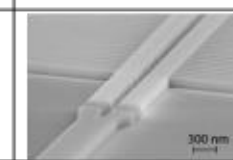 | 120 nm | $V_\pi L$ = 5 V mm, at 1kHz<br>$r_{33}$ = 30 pm/V at 1550 nm, | YLD124/APC (25wt%) | UW & Caltech |
| 2009 | 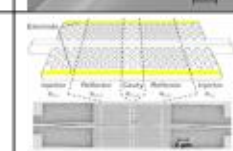 | 150 nm | $r_{33}$ = 9 pm/V at 1550 nm | AJ-CKL1-25wt% | Technische Universität Berlin |
| 2010 | 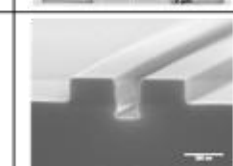 | 200 nm | $V_\pi L$ = 8 V mm<br>$r_{33}$ = 40 pm/V at 1550 nm | AJSP100 (15wt%), thin film $r_{33}$ of 65 pm/V | UW |
| 2010 | 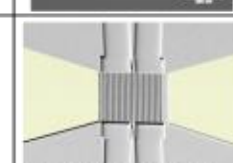 | 75nm | $V_\pi L$ = 0.56V mm<br>Effective $\gamma_{33}$ = 132 pm/V | AJ-CKL1-25wt% | UT-Austin<br>OO<br>UW |
| 2011 | 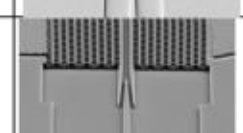 | 320nm | $V_\pi L$ = 0.44 V mm, at 100kHz<br>Effective $\gamma_{33}$ = 735 pm/V | AJ-CKL1-25wt% | UT-Austin<br>OO<br>UW |
| 2013 | 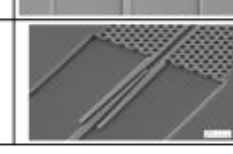 | 320nm | $V_\pi L$ = **0.345 V mm**, at 100kHz<br>Effective $\gamma_{33}$ = **1012 pm/V** | Soluxra's SEO125 25wt% | UT-Austin<br>OO<br>UW |

## V. Conventional Polymer Waveguide Modulator Fabricated by UV Imprinting and Ink-Jet Printing

The most common method for the fabrication of optical modulators and other nanophotonic devices includes using photolithography to define the pattern into a resist, and further transferring the pattern to the optical polymer via RIE. However, this method involves relatively complicated fabrication processes and low throughput. In this section, we will introduce a novel method combining imprinting and ink-jet printing techniques for the fabrication of polymeric optical modulators. Imprinting method is an effective method to achieve structural patterns with low cost and high fidelity [75], whereas ink-jet printing method provides great simplicity and flexibility in patterned feature deposition [4]. Both these methods are roll-to-roll compatible, thus possessing the potential for high-rate development of polymer based photonic devices [76].

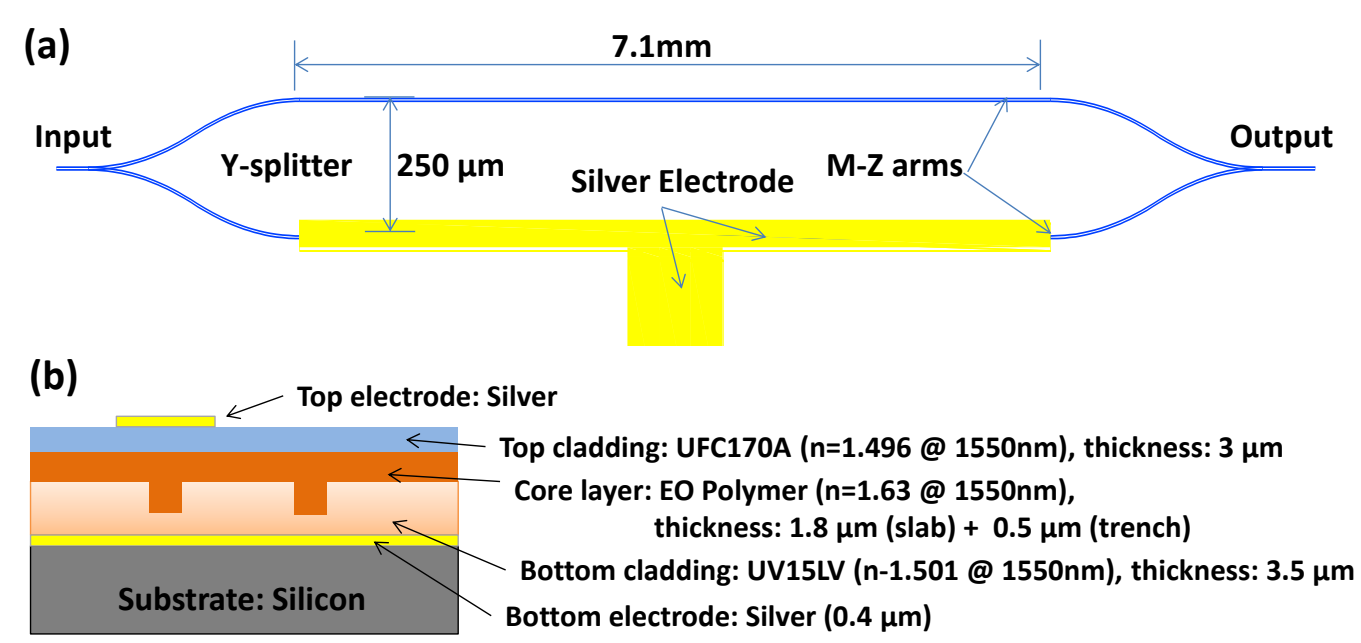

Fig. 13 (a). Schematic top view of the EO polymer MZI modulator. (b) Schematic cross section showing the different materials composing the EO polymer MZI modulator.

A schematic top view of our designed polymer MZI modulator is shown in Fig. 13 (a). A cross sectional schematic of the modulator is shown in Fig. 13 (b). The electrode separation and the arm length are designed to be d=8.3μm and L=7.1mm, respectively. The half-wave switching voltage is theoretically calculated to be $V_\pi = (\lambda \cdot d)/(L \cdot \gamma_{33}\, n^3) = 5.23V$.

In order to enable device development, the material system choice should meet certain criteria: 1) all the materials should satisfy the refractive index requirements to form an optical waveguide, 2) the bottom cladding layer should be imprintable, 3) the EO coefficient should be high enough to achieve index change with a small applied electric field, and 4) the core material should have suitable viscosity to be ink-jet printed. In order to satisfy the physical and chemical characteristics, we have selected UV15LV (n=1.50) as the bottom cladding layer, EO polymer (AJCKL1/APC, n=1.63) as the core layer, and UFC170A (n=1.49) as the top cladding layer. The choice of cladding materials is mainly based on the rigorous requirements of EO polymer because it is not compatible with solvent-based materials. UV15LV is a solvent-free polymer, and gets cross linked when exposed to UV. Besides, UV15LV is also an ink-jet printable material, thus having the potential to be deposited by ink-jet printing method in a roll-to-roll process. For the electrode layers, commercially available silver nanoparticle ink is chosen for ink-jet printing.

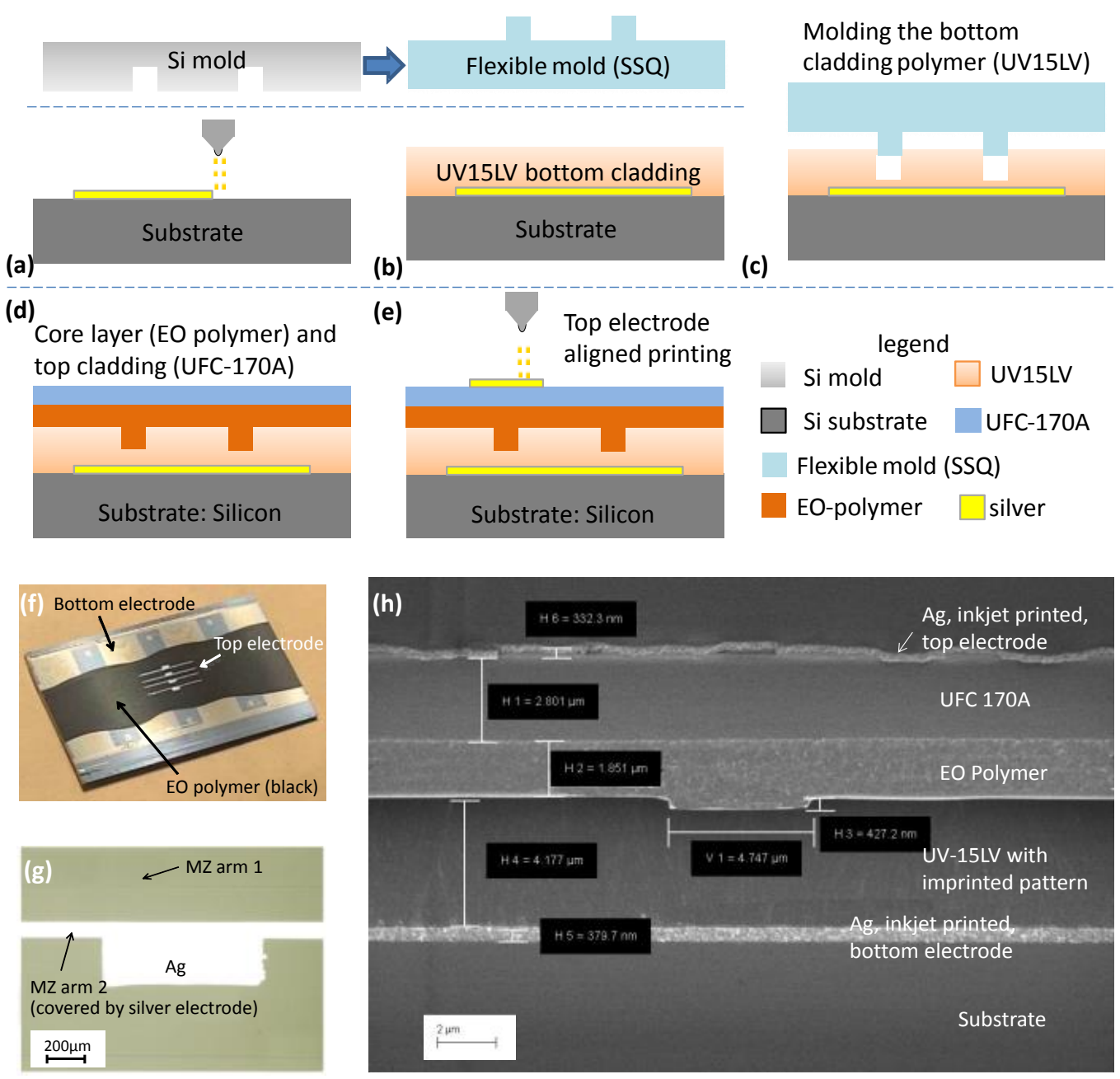

Fig.14. (a)-(e) Main process flow for fabricating an electro-optic polymer modulator by using imprinting and ink-jet printing method. (f) A fabricated modulator with ink-jet printed electrodes. (g) Microscopic image showing top silver electrode. (h) SEM image of device cross section.

The overall fabrication process flow is shown in Fig. 14. First, a 350-400nm silver ground electrode layer along with alignment marks on the substrate is ink-jet printed, as shown in Fig. 14 (a). Then, UV15LV is deposited onto the substrate to form bottom cladding layer, as shown in Fig. 14 (b). Next, a soft Epoxy Silsesquioxane (SSQ) mold containing the MZI structure is brought into conformal contact with the bottom cladding layer. Then, UV light is shone from the top to cure the UV15LV layer, followed by a demolding process, as shown in Fig. 14 (c). Upon completely curing the UV15LV layer, the EO polymer is coated, followed by top cladding UFC170A deposition, to form a trench waveguide structure, as shown in Fig. 14 (d). Finally, a top silver electrode is ink-jet printed on top of the top cladding layer, and aligned to one arm of the MZI waveguide [Fig. 14 (e)]. The ink-jet printed electrode serves as both a poling electrode and a driving electrode. Fig. 14 (f) shows a fabricated modulator with ink-jet printed top electrode. The microscope image of the top electrode and SEM image of the device cross section are shown in Fig. 14 (g) and (h), respectively. Then the device is poled by an electric field of 80V/μm at around 140 ℃.

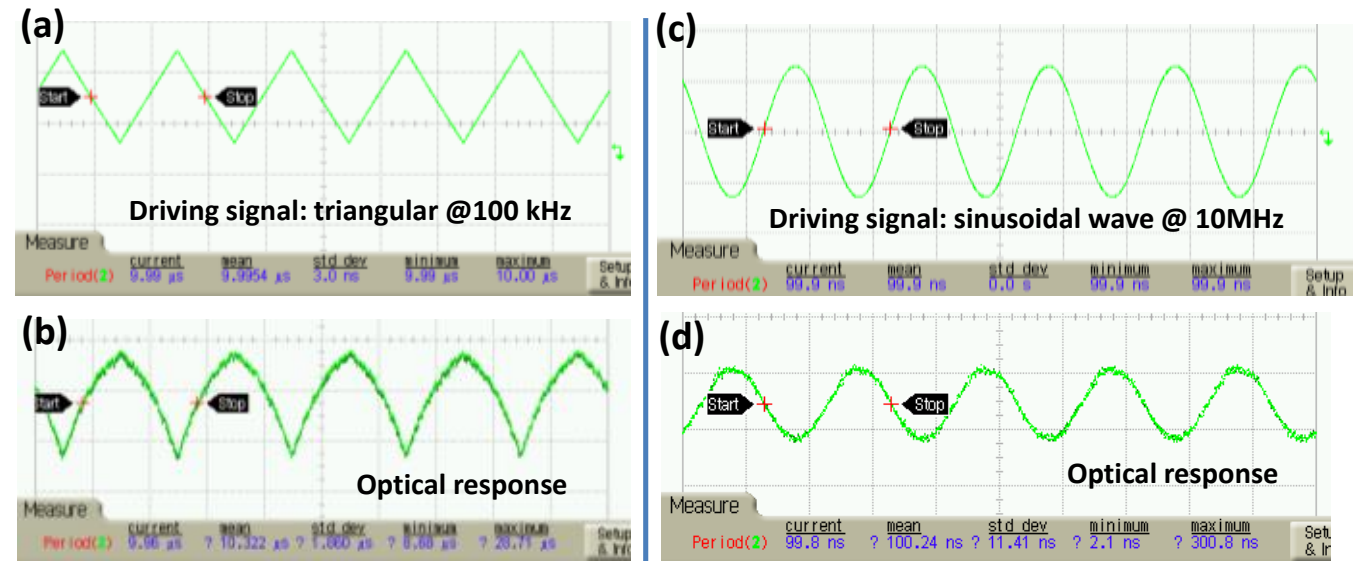

Fig. 15. (a-b) 100KHz triangular RF input and corresponding optical response. (c-d) 10MHz sinusoidal RF input and corresponding optical response.

To evaluate the modulation performance, the sample is mounted on an auto aligner for precise optical butt-coupling. TM-polarized light with 1550nm wavelength from a tunable laser is launched into the input waveguide through a polarization maintaining lensed fiber, and the output light is collected by a single mode lensed fiber. Driving RF signal is applied across the driving and ground electrodes of the device, and the modulated optical signal is then collected by a photodetector which is connected to an oscilloscope. Fig. 15 (a-b) shows the input RF triangular wave at 100KHz and the modulated optical signal at the same frequency. Fig. 15 (c-d) shows the input RF sinusoidal signal and the modulator response at 10MHz. By fine tuning the voltage applied to the point of over modulation, the $V_\pi$ is measured to be around 8.0V at 3KHz. It is higher than the calculated value probably due to the low poling efficiency of the EO polymer, which reduces the effective in-device $\gamma_{33}$ value to 52.3pm/V. This is the first demonstration of printed optical modulator to the best of our knowledge. Due to the utilization of lumped electrode which is not specially designed for high speed purpose, the modulation signal is weak at high frequencies. By utilizing traveling wave electrode [46] in our future work, we expect to further increase the operating speed of the device.

## VI. Polymer Waveguide Based Intra-Board and Inter-Board Optical Interconnects

In this section, we present our recent work on polymer based optical interconnects. There are two main categories of optical systems used in optical interconnects, namely free-space and guided-wave systems. In free-space optical interconnects, the optical field travels through a physically unconfined region between the source and destination. When applied in the inter-board scenario, it has the potential to offers high density, high speed, large fan-out capability and intrinsically low loss. As for intra-board scenario, the guided-wave method is more practical. Waveguides or optical fibers are used to form the optical path [77]. This provides ease of packaging, flexibility of route design, and system reliability. The optical layer can be embedded in the middle of the PCB board or placed on the surface. Here, we have demonstrated a few examples from intra-board to inter-board optical interconnects based on polymers.

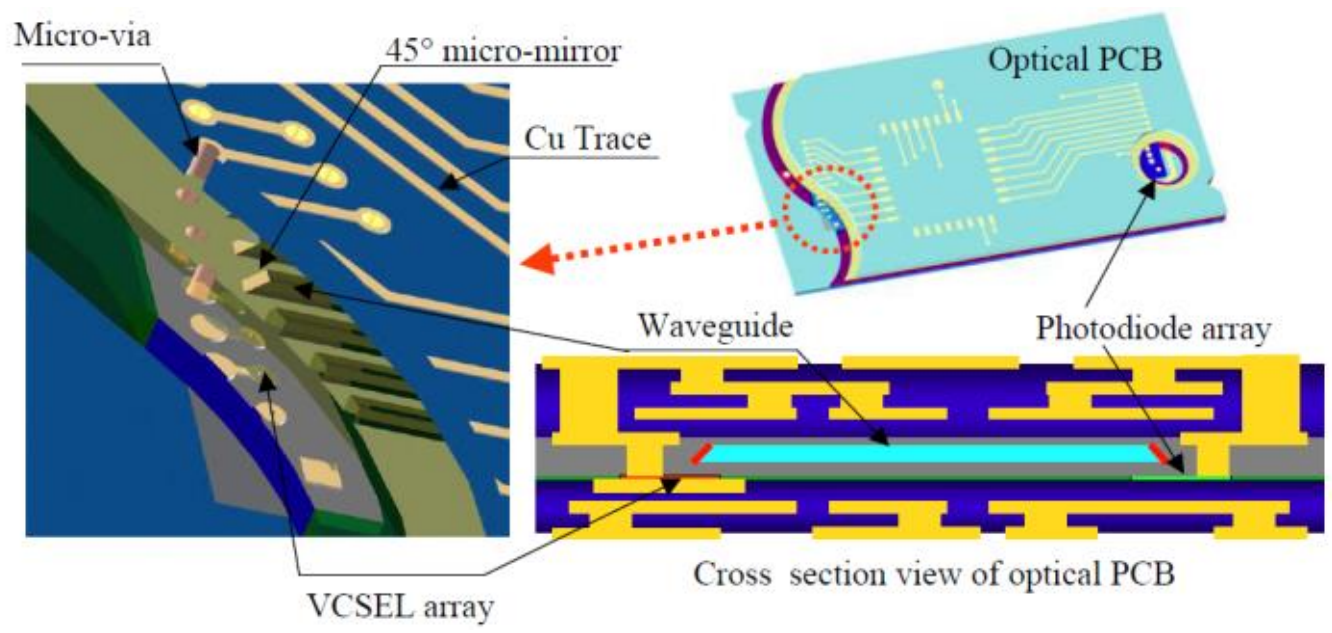

Fig. 16. Schematic of the side view of the vertical integration layers. Waveguides, VCSEL, photodetector, waveguide coupler, electrical vias, and other electrical interconnection layers are clearly shown.

For fully embedded board-level or intra-board guided-wave optical interconnects [78], the topology is shown in Fig.16, in which all elements involved in providing high-speed optical communications within one board are shown. As can be seen, main components include a pair of vertical-cavity surface-emitting laser (VCSEL) and photo detector (PD), 45° mirrors as surface-normal couplers, and a polyimide-based channel waveguide functioning as the physical layer of optical bus. The driving electrical signal to modulate the VCSEL, and the demodulated signal received at the photo detector are all routed through electrical vias to the surface of the PC board. In addition, the vias also function as heat dissipation paths for VCSEL and PD.

When placing the optical layer on the surface of a PCB board, the layer can provide additional functionalities by not only realizing intra-board optical interconnects, but also providing the interface to communicate with layers on other boards. Using molding on the top surface of a PCB board, we have demonstrated an array of 12 point-to-point waveguides [79] and a 3-to-3 bi-directional optical bus architecture [80] based on multimode polymer waveguides with embedded 45° mirrors. Furthermore, by placing two boards back-to-back, short-distance free space optical interconnects are realized with the help of proximity surface normal couplers.

The main concept is illustrated in Fig. 17 (a) which contains two optical links using 50 μm×50 μm polymer waveguides and proximity couplers. The 1st link from board 1 to board 2 contains 4 mirrors and the 2nd link from board 1 to board 3 contains 2 mirrors. In order to reduce the beam divergence and make coupling more effective, we employed ink-jet printing technology to precisely place microlenses [profile shown at the lower left of Fig. 17 (a)] right on top of the reflective mirrors. From the measurement results, the propagation loss of the polymer waveguide is found to be 0.18dB/cm, and each 45° mirror contributes 1.9dB loss. By comparing the situation with and without microlenses in between, we estimate that each integrated microlens can provide 1.5dB improvement at shorter board-to-board separation (~1-2mm) and 3.7dB improvement at larger separation (>4mm). Using the optical link from board 1 to board 3, we can perform high speed data communication test.

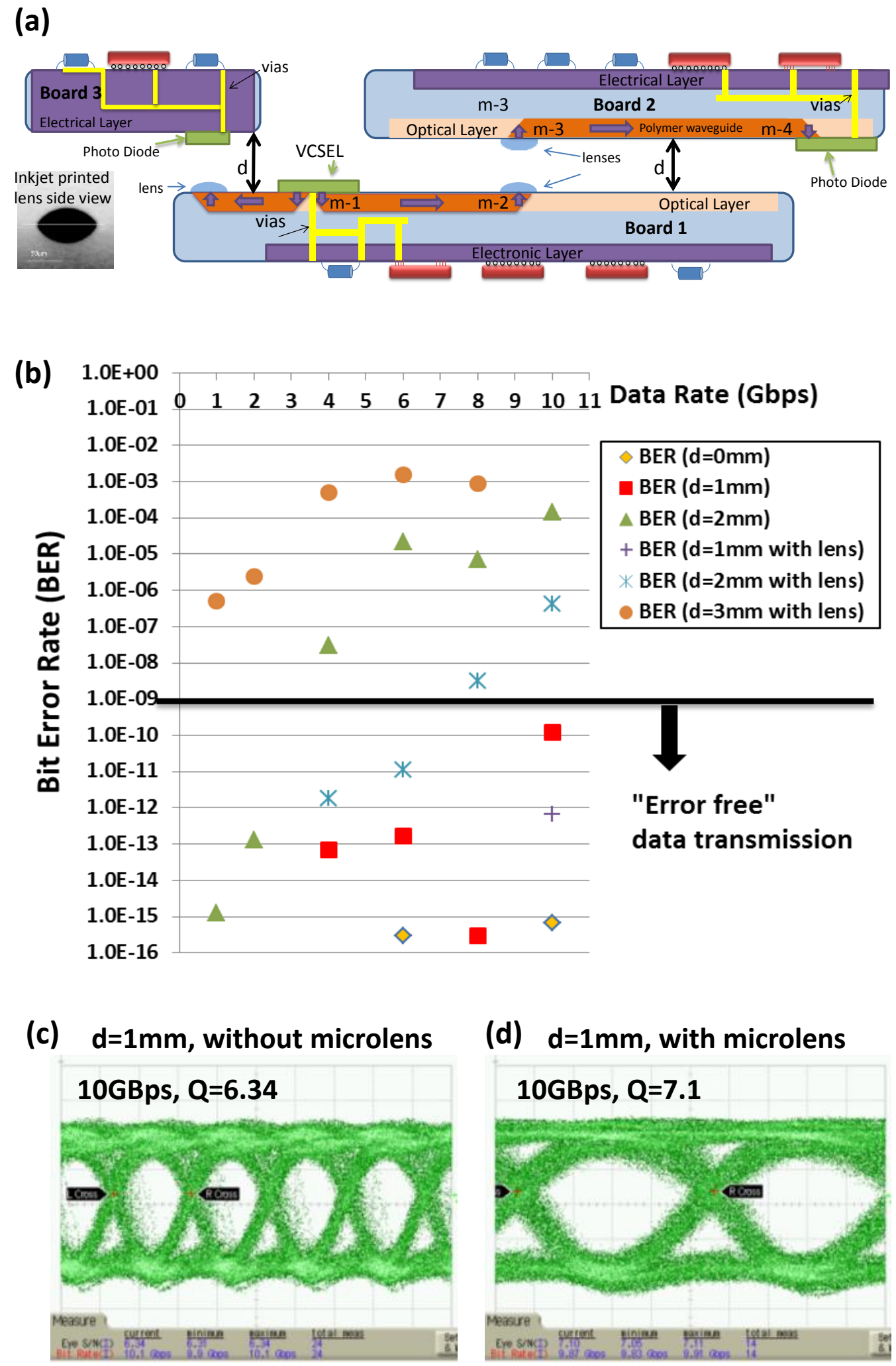

Fig. 17 (a) The schematic of inter- and intra-board optical interconnects with polymer waveguides and 45° mirrors with ink-jet printed microlenses. The profile of ink-jet printed lens with 70 μm diameter is shown at the lower left. (b) Bit Error Rate (BER) distribution with data rate at different separations with/without ink-jet printed lens. (c) Eye diagram showing 10Gbps data transmission when board separation is 1mm. (d) Eye diagram showing improved 10Gbps data transmission quality when board-to-board separation is 1mm, with a microlens inserted in the optical path.

1550nm light is coupled vertically in/out the waveguide of board 1 by two embedded 45° mirrors and collected by photo diode located at board 3, which is at the separation of d to board 1. The experimental results shown in Fig. 17 (b) reveal that at separation less than 1mm the proximity coupler can easily handle error-free data transmission, which is defined as the data transmission with bit error rate (BER) $<10^{-9}$, at 10Gbps. In addition, it can be seen in Fig. 17 (b) that, as the separation increases, the transmission quality decreases drastically due to free space beam divergence. Without a microlens, at separation of 2mm, only data rate below 3Gbps can be transmitted error-free. On the other hand, at the same separation, if one microlens is integrated to reduce beam divergence, it can support error-free data transmission up to 7Gbps with improved signal quality. The eye diagrams of the 10Gbps transmission at 1mm separation, with and without ink-jet printed microlens, are shown in Fig.17 (c-d), respectively, from which the improvement of the signal quality with the help of the microlens can be seen. To the best of our knowledge, this is the first report of free-space coupling between waveguides on separate boards.

## VII. Conclusion

We reported our recent work on several polymer based optical modulators for on-chip and board-level photonic applications. The highly linear broadband directional coupler modulator was demonstrated with bandwidth-length product of 125GHz·cm, the 3-dB electrical bandwidth of 10GHz, and the SFDR of 110±3dB/Hz$^{2/3}$ over 2-8GHz, which benefited from the $\Delta\beta$-reversals and traveling wave electrode. To the best of our knowledge, such high linearity is first measured at a frequency up to 8GHz by our group.

A 75nm-wide slot PCW modulator refilled with EO polymer was demonstrated to have 23dB modulation enhancement due to slow light effect, and a low $V_\pi \times L$ of 0.56V mm. The band-engineered EO polymer refilled 320nm-wide slot PCW achieved the effective in-device $\gamma_{33}$ of 1012pm/V and $V_\pi \times L$ of 0.345V mm, which is the best figure of merit of nanophotonic modulators to our knowledge.

Finally, imprinted and ink-jet printed active (modulator) and passive (board-to-board coupler) devices were presented. A fully functional EO polymer MZI modulator was developed and demonstrated to work at $V_\pi$=8V. A 45° mirror with integrated microlens was shown to enable free-space interconnection between waveguides, VCSELs, and photodetectors on separate boards, and demonstrated to work at over 10Gbps with BER better than $10^{-9}$. The fabrication processes involved are fully roll-to-roll compatible, which can enable high throughput, low cost and volume manufacturing of photonic devices.

## Acknowledgment

We would like to acknowledge the contributions from previous team members in this invited review article. Their names can be found in the references. Technical advice provided by AFRL, AFOSR and DARPA program managers including Drs. Rob Nelson, Charles Lee, Gernot Pomrenke and Dev Shenoy, are greatly appreciated.

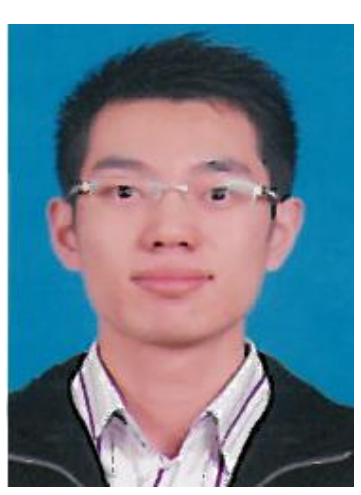

**Xingyu Zhang,** received his B.S. degree in Electrical Engineering from Beijing Institute of Technology, Beijing, China, in 2009, and the M.S. degree in Electrical Engineering from University of Michigan, Ann Arbor, MI, in 2010. He is currently a P.hD student in University of Texas, Austin, TX. His current research focuses on the design, fabrication, and characterization of silicon and polymer nanophotonic devices, including highly linear broadband optical modulator, slotted photonic crystal waveguide modulator, and electro-magnetic field sensor. His previous work includes gyroscopic optical forces and hybrid diode-microresonator lasers.

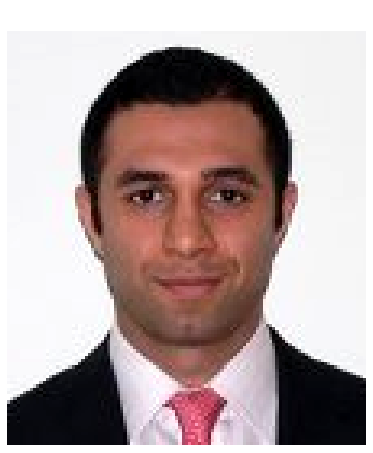

**Amir Hosseini** (S'05–M'13), received his B.Sc. degree in Electrical Engineering in 2005 from Sharif University of Technology, Tehran, Iran, M.Sc. degree in Electrical and Computer Engineering in 2007 from Rice University, Houston, TX, and the Ph.D. degree in Electrical and Computer Engineering from the University of Texas at Austin, Texas in 2011.

He has been engaged in research on modeling, design, fabrication and characterization of optical phased array technology, true-time delay lines, and high performance optical modulators. He is a Prince of Wales' scholar in 2011 and recipient of the Ben Streetman Award in 2012, and has authored or co-authored over 70 peer reviewed technical papers. He is a member of IEEE, OSA, and SPIE. He has been serving as the principal investigator for an AFRL sponsored project on polymer optical modulators since 2012.

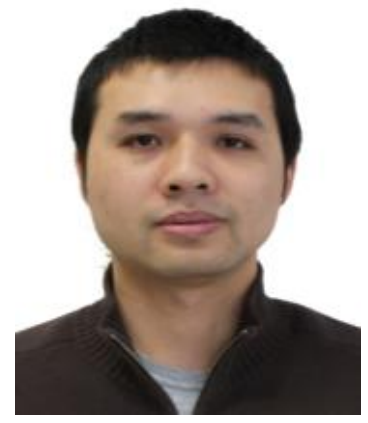

**Xiaohui Lin,** received his B.S and Ph.D degrees in mechanical engineering from the Huazhong University of Science and Technology, Wuhan, China, in 2004 and 2008, respectively. His research interests include MEMS and polymer based nanophotonic devices for optical interconnects. He is currently working on applying imprinting and ink-jet printing techniques for low cost electrical and optical devices.

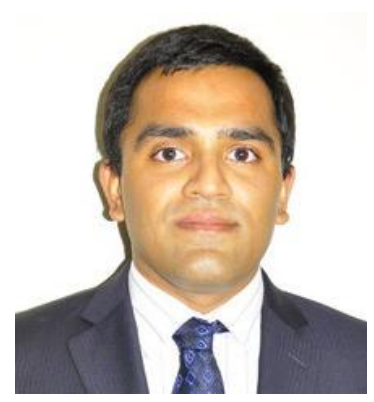

**Harish Subbaraman** (M'09), received his M.S. and Ph.D. degrees in Electrical Engineering from the University of Texas at Austin, Texas in 2006 and 2009, respectively. With a strong background in RF photonics and X-band Phased Array Antennas, Dr. Subbaraman has been working on optical true-time-delay feed networks for phased array antennas for the past 7 years. Throughout these years, he has laid a solid foundation in both theory and experimental skills. His current research areas include printing and silicon nanomembrane based flexible electronic and photonic devices, polymer photonics, slow-light photonic crystal waveguides, carbon nanotube and silicon nanoparticle nanofilm based ink-jet printed flexible electronics, and RF photonics; He has served as a PI on 7 SBIR/STTR Phase I/II projects from NASA, Air Force and Navy. Dr. Subbaraman has over 50 publications in refereed journals and conferences.

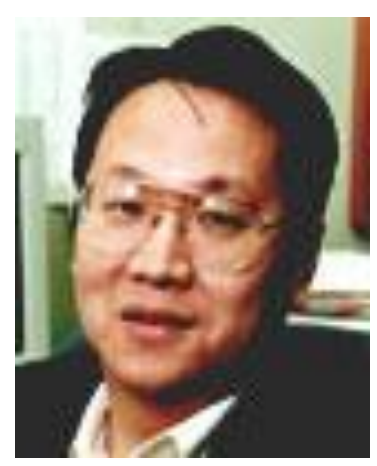

**Ray T. Chen** (M'91–SM'98–F'04) holds the Cullen Trust for Higher Education Endowed Professorship at UT Austin and the director of nanophotonics and optical interconnects research lab within the microelectronics research center. He is also the director of a newly formed AFOSR MURI-Center involved faculty from Stanford, UIUC, Rutgers for Silicon Nanomembrane. He received his BS degree in Physics from National Tsing-Hua University in 1980 in Taiwan and his MS degree in physics in 1983 and his PhD degree in Electrical Engineering in 1988, both from the University of California. He joined UT Austin as a faculty to start optical interconnect research program in the ECE Department in 1992. Prior to his UT's professorship, Chen was working as a research scientist, manager and director of the Department of Electrooptic Engineering in Physical Optics Corporation in Torrance, California from 1988 to 1992.

Chen also served as the CTO/founder and chairman of the board of Radiant Research from 2000 to 2001 where he raised 18 million dollars A-Round funding to commercialize polymer-based photonic devices. He also serves as the founder and Chairman of the board of Omega Optics Inc. since its initiation in 2001. Over 5 million dollars of research funds were raised for Omega Optics. His research work has been awarded with 110 research grants and contracts from such sponsors as DOD, NSF, DOE, NASA, NIH, EPA, the State of Texas, and private industry. The research topics are focused on three main subjects: 1. Nano-photonic passive and active devices for optical interconnect and bio-sensing applications, 2. Polymer-based guided-wave optical interconnection and packaging, and 3. True time delay (TTD) wide band phased array antenna (PAA). Experiences garnered through these programs in polymeric material processing and device integration are pivotal elements for the research work conducted by Chen's group.

Chen's group at UT Austin has reported its research findings in more than 650 published papers including over 85 invited papers. He holds 20 issued patents. He has chaired or been a program-committee member for more than 90 domestic and international conferences organized by IEEE, SPIE (The International Society of Optical Engineering), OSA, and PSC. He has served as an editor, co-editor or coauthor for 22 books. Chen has also served as a consultant for various federal agencies and private companies and delivered numerous invited talks to professional societies. Dr. Chen is a Fellow of IEEE, OSA and SPIE. He was the recipient 1987 UC Regent's dissertation fellowship and of 1999 UT Engineering Foundation Faculty Award for his contributions in research, teaching and services. He received IEEE Teaching Award in 2008. Back to his undergraduate years in National Tsing-Hua University, he led a university debate team in 1979 which received the national championship of national debate contest in Taiwan. 44 students have received the EE PhD degree in Chen’s research group at UT Austin.